\documentclass{jfm}

\usepackage{graphicx,afterpage,hhline,url}
\usepackage{natbib}
\usepackage{amsmath,amstext,amssymb,upmath}
\usepackage{subfigure}

\ifCUPmtlplainloaded \else
  \checkfont{eurm10}
  \iffontfound
    \IfFileExists{upmath.sty}
      {\typeout{^^JFound AMS Euler Roman fonts on the system,
                   using the 'upmath' package.^^J}%
       \usepackage{upmath}}
      {\typeout{^^JFound AMS Euler Roman fonts on the system, but you
                   dont seem to have the}%
       \typeout{'upmath' package installed. JFM.cls can take advantage
                 of these fonts,^^Jif you use 'upmath' package.^^J}%
      }
  \else
  \fi
\fi

\ifCUPmtlplainloaded \else
  \checkfont{msam10}
  \iffontfound
    \IfFileExists{amssymb.sty}
      {\typeout{^^JFound AMS Symbol fonts on the system, using the
                'amssymb' package.^^J}%
       \usepackage{amssymb}%
         \let\leq=\leqslant
         \let\geq=\geqslant
      }{}
  \fi
\fi

\ifCUPmtlplainloaded \else
  \IfFileExists{amsbsy.sty}
    {\typeout{^^JFound the 'amsbsy' package on the system, using it.^^J}%
     \usepackage{amsbsy}}
    {\providecommand\boldsymbol[1]{\mbox{\boldmath $##1$}}}
\fi

\newcommand{\Rd}{R}
\newcommand{\lonset}{\ell}

\title[Faraday instability on a sphere: numerical simulation]{Faraday instability on a sphere: \\numerical simulation}

\author[A. Ebo-Adou, L.S.~Tuckerman, S.~Shin, J.~Chergui and D.~Juric]%
       {A. Ebo-Adou$^{1,2,3}$, L.S.~Tuckerman$^1$
\thanks{Email address for correspondence: laurette@pmmh.espci.fr},\ns S.~Shin$^4$, J.~Chergui$^2$ and D.~Juric$^2$}%

\affiliation{
$^1$Physique et M\'ecanique des Milieux 
H\'et\'erog\`enes (PMMH),
CNRS, ESPCI Paris, PSL Research University, Sorbonne Universit\'e, 
Univ. Paris Diderot, 75005 France
\\[\affilskip]
  $^2$Laboratoire d'Informatique pour la M\'ecanique et les Sciences de l'Ing\'enieur (LIMSI),
Centre National de la Recherche Scientifique (CNRS), Universit\'e Paris Saclay,
B\^at. 507, Rue du Belv\'ed\`ere, Campus Universitaire, 91405 Orsay, France
\\[\affilskip]
$^3$
{Institut des Sciences de la Terre, Centre d'\'Etudes et de Recherche de Djibouti, Route de l'a\'eroport B.P 486 Djibouti-ville, R\'epublique de Djibouti}
\\[\affilskip]
$^4$Department of Mechanical and System Design Engineering, Hongik University, Seoul 121-791, Republic of Korea}
\begin{document}

\maketitle

\begin{abstract}
  We consider a spherical variant of the Faraday problem, in which a
  spherical drop is subjected to a time-periodic body force, as well
  as surface tension.
  We use a full three-dimensional parallel front-tracking code to
  calculate the interface motion of the parametrically forced
  oscillating viscous drop, as well as the velocity field inside and
  outside the drop. Forcing frequencies are chosen so as to excite
  spherical harmonic wavenumbers ranging from 1 to 6.  We excite
  gravity waves for wavenumbers 1 and 2 and observe translational and
  oblate-prolate oscillation, respectively.  For wavenumbers 3 to 6,
  we excite capillary waves and observe patterns analogous to the
  Platonic solids.  For low viscosity, both subharmonic and
  harmonic responses are accessible.  The patterns arising in each
  case are interpreted in the context of the theory of pattern
  formation with spherical symmetry.
\end{abstract}
{\it published in Journal of Fluid Mechanics {\bf 870}, 433-459 (2019)}

\section{Introduction}

The \cite{Far1831} instability, in which the
interface between two superposed fluid layers subjected to periodic
vertical vibration of sufficient amplitude forms sustained standing
wave patterns, has been instrumental in the study of pattern formation,
leading to the discovery and analysis of hydrodynamic quasipatterns
\citep{EF1994,Rucklidge2015},
superlattices \citep{KPG1998,Silber1998,Arbell2002},
supersquares \citep{Douady1990,Lyes} and other exotic
patterns \citep{Perinet2012,Rajchenbach2011}.

Here, we consider a spherical analogue to the Faraday instability,
a fluid drop subjected to a time-periodic radial body force.
The study of such problems therefore
relies particularly on numerical simulation.

Numerical studies of axisymmetric oscillating viscous drops have been
carried out by \cite{LM1988} via a boundary-integral
method, by \cite{Patz1991} and \cite{Meradji2001} via a
Galerkin/finite-element method, and by \cite{Bas1992} via a
marker-in-cell initial-value problem. 
\cite{TB1983} used a Poincare-Lindstedt expansion method, 
to calculate the shapes of axisymmetric inviscid drops
subjected to moderate amplitude oscillations for the three lowest
capillary modes (see \eqref{eq:defint_intro}).
They were unable to confirm mathematically the existence of an
asymptotic finite-amplitude motion; this can be verified only
by means of numerical calculations. Indeed, the problem of 
axisymmetric ellipsoidal drop oscillation and decay has now come to be
seen as a routine validation test case for numerical codes and
interface methods for multiphase flows.  However, the full non-linear
problem is non-axisymmetric and requires three-dimensional numerical
simulation with interface algorithms that ensure volume conservation
and precise calculation of capillary forces as well as the ability to
integrate highly spatially resolved systems over long physical times.

Recent advances have led to powerful general purpose codes such as
Gerris \citep{Popinet2003} and BLUE
\citep{shin2017solver}.
It is with the multiphase code BLUE, which is based
on recently developed hybrid front-tracking/level-set interface methods implemented on
parallel computer architectures, that we conduct the current study.
We have previously used this code to study large-scale square patterns
of Faraday waves \citep{Lyes}.  Here, we use BLUE to carry out the 
first numerical investigation of the Faraday problem on a sphere.  

One of the most appealing aspects of the Faraday instability
is that the pattern length scale is not set by the geometry,
but by the imposed forcing frequency.
(It is this feature which has allowed the generation of
quasipatterns and superlattices, since multiple length
scales can be excited simultaneously over the entire domain
by superposing different frequencies.)
This means that for a drop with a fixed radius $\Rd$,
patterns can be created with any wavenumber $\ell$,
where the interface is described via its spherical harmonic decomposition
\begin{equation}
\zeta(t,\theta,\phi) = \sum_{\ell=0}^\infty \sum_{m=-\ell}^{m=\ell} \zeta_\ell^m(t)Y_\ell^m(\theta,\phi)
\label{eq:defint_intro}
\end{equation}
For fixed $\Rd$, the length scale associated with $\ell$ is $\Rd/\ell$.
More importantly, $\ell$ is associated with a set of allowed patterns;
each value of $\ell$ leads to a qualitatively different situation.  We
will explore the motion and shape of an oscillating drop 
for values of $\ell$ up to 6.

\section{Methods}
\subsection{Pattern formation on a sphere}

Our previous paper \citep{Ali1} concerns the linear stability, via
Floquet analysis, of the spherical Faraday problem.
As is the case for all spherically symmetric problems, 
the equation 
governing the linear stability 
does not depend on the order $m$ of the spherical harmonic.
Therefore, a bifurcation from the spherically symmetric state involves
$2\ell+1$ linearly independent solutions ($Y_\ell^m$ with $ -\ell \leq
m \leq \ell$) with the same growth rate.  The combination of these
modes, i.e. the pattern, that can result from such a bifurcation is
determined by the nonlinear terms.
A pattern with a given $\ell$ cannot be associated with a unique
combination of modes $m$, since rotation of a spherical harmonic
changes $m$ (but not $\ell$).


Symmetry groups provide a classification of patterns which does not
depend on orientation. A number of researchers 
\citep{Busse1975,BR1982,Riahi1984,IG1984,Golubitsky,Chossat1991,Matth2003}
have studied the patterns which are allowed and those which are
preferred for various values of $\ell$.  
Patterns that appear at bifurcations can be associated with 
a subgroup of the group $O(3)$ of symmetries of the sphere.
The subgroups of interest  are $O(2)$, $D_k$, and the
exceptional subgroups
$T$, $O$, and $I$.
$O(2)$ consists of the symmetries of a circle; patterns which
are axisymmetric (for some orientation) are in this category.
$D_k$ describes the symmetries of a $k$-gon, and thus patterns which
are invariant under reflection and rotation by $2\pi/k$ (about
a fixed axis). The three exceptional subgroups are associated with
the five Platonic solids: $T$ (tetrahedron), $O$ (octahedron or cube), and
$I$ (icosahedron or dodecahedron).

The Platonic solids are regular polyhedra.  Although a drop does not
have angular vertices and flat faces, a polyhedron with the same
symmetry properties
can be constructed from a drop in a $T$, $O$, or $I$ configuration
by assigning a local maximum on the surface to a
vertex and a local minimum to a face.  The dual of a polyhedron is
obtained by inverting its vertices with its faces, an operation which
preserves symmetry.  Similarly, the dual of a drop can be formed by
inverting maxima and minima.  An oscillating drop provides an ideal
opportunity to observe duality: since a location which is the site of
a maximum contains a minimum after half of an oscillation period, the
interface alternates between a pattern and its dual.  For the Platonic
solids, the dual of an octahedron is a cube, that of an icosahedron is
a dodecahedron and the dual of a tetrahedron is another tetrahedron.

The patterns are highly dependent on the value of $\ell$ considered.
Axisymmetric (sometimes called zonal) solutions are never stable if
$\ell\geq 2$ \citep{Chossat1991}.  For odd $\ell$, solutions exist
which are stable at onset.  In contrast, for even $\ell$, all
solutions are unstable near the bifurcation point.  In this case, the
preferred solution can be considered to be that with the smallest number
of unstable eigenvalues \citep{Matth2003} or which extremizes a functional
\citep{Busse1975}. Unstable solution branches
produced at a transcritical bifurcation can be stabilized, for example
at a saddle-node bifurcation some distance from the threshold.
%

The theory of pattern selection with $O(3)$ symmetry
differs from the framework of our simulations in some important ways. 
First, the theory applies to steady bifurcations rather than oscillatory
solutions resulting from time-periodic forcing. 
The archetypical application is Rayleigh-B\'enard convection
in a sphere \citep{Busse1975,BR1982,Riahi1984}
and the steady symmetry-breaking bifurcations it undergoes.
However, many of the conclusions
can be generalized to the time-periodic context,
by considering the discrete-time dynamical system derived
by sampling the continuous-time system at a single
phase of the forcing period.


Secondly, the theory applies close to the threshold.  Our simulations
are carried out far from threshold, so that instabilities can grow on
a reasonable timescale and stabilize at an amplitude that can be
clearly seen.
Our patterns contain modes generated by nonlinear interactions
and their existence or
stability may result from secondary bifurcations.

Despite these differences, we will see that there is much 
common ground between the patterns we observe and those predicted
by theory.

\subsection{Problem formulation, governing equations and numerical scheme}
\label{sec:numerics}

The governing equations for an incompressible two-phase flow can be
expressed by a single field formulation:
\begin{equation}\label{NS-Eqs}
\displaystyle{\rho  \left( \frac{\partial \textbf{u}}{\partial t}
+\textbf{u}\cdot\nabla \textbf{u}\right) =  -\nabla P + \rho \textbf{G}
+ \nabla \cdot \mu\left(\nabla\textbf{u} +\nabla\textbf{u}^T \right)
+ \textbf{F}}, \qquad \displaystyle{\nabla\cdot\textbf{u}=0 }
\end{equation}
where $\textbf{u}$ is the velocity, $P$ is the pressure, $\rho$ is the
density, $\mu$ is the dynamic viscosity and $\textbf{F}$ is the local surface
tension force at the interface. Here, ${\bf G}$ is 
an imposed time-dependent radial acceleration:
\begin{equation}
{\bf G}=-\left(g+ a \cos(\omega t)\right)\frac{r}{R}\textbf{e}_r
\label{eq:grav}\end{equation}
where $g$ is a constant acceleration, referred to for simplicity as
gravitational, which is set to zero when we carry out capillary simulations.
$\Rd$ is the radius of the drop, $\textbf{e}_r$ is the radial unit
vector, and $a$ and $\omega$ are the amplitude and frequency of the
oscillatory forcing.

Material properties such as density or viscosity are defined in the
entire domain:
\begin{equation}
\left.\begin{array}{c}
\rho(\textbf{x},t) = \rho_{1} +\left(  \rho_{2} -\rho_{1}\right)I(\textbf{x},t)\\
\mu(\textbf{x},t) = \mu_{1} +\left( \mu_{2} -\mu_{1}\right)I(\textbf{x},t).
\end{array}\right.
\label{eq:rhomu}\end{equation}
The indicator function $I$ in \eqref{eq:rhomu} is the Heaviside function, 
whose value is zero in one phase and one in the other
phase.  In our discrete numerical implementation $I$
is approximated by $I_{\rm num}$, which represents
a smooth transition across 3 to 4 grid cells,
as pioneered in the immersed boundary method of \cite{Pes1977}.  In
our method $I_{\rm num}$ is generated using a vector distance function computed
directly from the tracked interface \citep{ShJu09}.

The fluid variables $\textbf{u}$ and $P$ are calculated by a
projection method 
\citep{Chorin-mcomp-1968}. 
The temporal scheme is first order, 
with implicit time integration used for the viscous terms.
For spatial discretization we use the staggered-mesh
marker-in-cell (MAC) method
\citep{HW1965}
on a uniform finite-difference grid 
with second-order essentially non-oscillatory (ENO) advection
\citep{SO1989}.
The pressure and
distance function are located at cell centers while the $x$, $y$ and
$z$ components of velocity are located at the faces. All spatial
derivatives are approximated by standard second-order centered
differences. The treatment of the free surface uses a hybrid
Front-Tracking/Level-Set technique which defines the interface both by
the Level-Set distance function field
on the Eulerian grid as well as by 
triangles on the Lagrangian interface mesh.

The surface tension $\textbf{F}$ is implemented by the hybrid/compact formulation
\citep{Shin07}
\begin{equation}
\textbf{F} = \sigma \kappa_{H} \nabla I , \qquad
\kappa_{H}=\frac{\textbf{F}_{L}\cdot\textbf{N}}{\textbf{N}\cdot\textbf{N}}\end{equation}
where $\sigma$ is the surface tension coefficient and $\kappa_{H}$ is
twice the mean interface curvature field calculated on the Eulerian
grid, with
\begin{equation}
\textbf{F}_{_{L}} = \int_{\Gamma(t)} \kappa_f  \textbf{n}_{f} \delta_f
\left( \textbf{x} - \textbf{x}_{f} \right) \mathrm{d}a ,
\qquad \textbf{N} = \int_{\Gamma(t)}  \textbf{n}_{f} \delta_f
\left( \textbf{x} - \textbf{x}_{f} \right) \mathrm{d}a
\end{equation}
Here, $\textbf{x}_f$ is a parameterization of the time-dependent
interface, $\Gamma(t)$, and
$\delta_f \left( \textbf{x} -\textbf{x}_{f} \right) $
is a Dirac distribution that is non-zero only
where $\textbf{x}=\textbf{x}_f$; $\textbf{n}_f$ stands for the unit
normal vector to the interface and $ \mathrm{d}a$ is the area of an
interface element; $\kappa_f$ is twice the mean interface curvature
obtained on the Lagrangian interface. The geometric information, unit
normal, $\textbf{n}_f$, and interface element length, $ \mathrm{d}a$
in $\textbf{N}$ are computed directly from the Lagrangian interface
and then distributed onto the Eulerian grid using the discrete delta
function and the immersed boundary method of
\cite{Pes1977}.
A detailed description of the procedure for calculating $\textbf{F}$,
$\textbf{N}$ and $I_{\rm num}$ can be found in
\cite{ShJu07,ShJu09},
where in particular we demonstrate that this method of calculating the surface tension force reduces any parasitic currents in the standard static drop test case to a level of
$O(10^{-7})$ for fluids with properties similar to those we use here.

The Lagrangian interface is advected by integrating
$\mathrm{d}\textbf{x}_f/\mathrm{d}t = \textbf{V}$ with a
second-order Runge-Kutta method where the interface velocity,
$\textbf{V}$, is interpolated from the Eulerian velocity.

The parallelization of the code is based on algebraic domain
decomposition, where the velocity field is solved by a parallel 
generalized minimum residual (GMRES)
method for the implicit viscous terms and the pressure by a parallel
multigrid method motivated by the algorithm of
\cite{Kwak-InterScience-2004}.  Communication across process threads
is handled by message passing interface (MPI) procedures.

The code contains a module for the definition of immersed solid
objects and their interaction with the flow, which we have used to
simulate Faraday waves in a spherical container. In order
to simulate a fluid within a solid sphere we take the simple approach
of defining all grid cells whose centers lie within the solid region
as solid. Then Dirichlet (no-slip) boundary conditions for the velocity
and Neumann conditions for the pressure are 
applied to those cell faces as in the projection method.
On a Cartesian grid this
necessarily creates a stair-stepped solid/fluid boundary; however it
is found that the method works well in practice since the discrete
momentum flux is conserved and is simpler than
and equivalent to other approaches which impose a near-wall force to
ensure a no-slip condition at the solid \citep{Trygg2011}.
%
%
See \cite{shin2017solver} for further details.

\subsection{Physical parameters of the fluids}
\label{sec:phys_par}

Our numerical code
can treat inner and outer spherical domains of any size containing fluids
of any density $\rho$, viscosity $\nu$, and surface tension $\sigma$, 
leading potentially to a large number of non-dimensional parameters, in
addition to those describing the forcing amplitude $a$ and frequency $\omega$.
We have chosen to limit the parameter space as follows.

The density and viscosity ratios that we have chosen are
typical of oil droplets in air.
The inner and outer densities are
$\rho_d=940$~kg/m$^3$ and $\rho_{\rm out}=1.205$~kg/m$^3$,
leading to the density ratio $\rho_d/\rho_{\rm out}=780$, 
while the inner and outer kinematic viscosities are 
$\nu_d = 10^{-5}$~m$^2$/s
and $\nu_{\rm out}=1.5\times 10^{-5}$~m$^2$/s, 
leading to the viscosity ratio of $\nu_d/\nu_{\rm out}=0.66$.
The viscosity of both fluids is sufficiently low that the stability
diagram approaches that of the Mathieu equation.
(One of the conclusions of \cite{Ali1} is that the Mathieu equation
describes the inviscid Faraday instability even in a spherical
geometry.)  
More specifically, $\nu_d/(\Rd^2\omega)$ is between
$3\times 10^{-4}$ and $7\times 10^{-3}$ for the capillary cases and between
$6\times 10^{-4}$ and $9\times 10^{-4}$ for the gravitational cases.
Usually, Faraday waves are subharmonic, i.e. their period is twice
the forcing period $T$. However, in this low-viscosity regime, 
we can easily excite harmonic waves as well \citep{K1996}, whose period is
the same as the forcing period.

For the capillary cases, we set the surface tension to be
$\sigma=0.02$ kg/s$^2$, the constant radial acceleration to be $g=0$,
and the drop radius to be $R=0.06$ m, while for the gravitational
cases, we set the constant radial acceleration to be $g=-1$ m/s$^2$,
the surface tension to be $\sigma=0$, and the drop radius to be
$R=0.05$ m. We will use these parameters to non-dimensionalize
the forcing frequencies and amplitudes as in equation \eqref{eq:nondim}
and list these in section \ref{sec:cases} and table \ref{tab:forcing_response}.

\subsection{Numerical parameters}

For the capillary cases, the domain is a sphere whose radius is $2\Rd$ (see section \ref{sec:numerics}
and figures \ref{fig:l=2} and \ref{fig:l=2_out}).
The Eulerian mesh is uniform and Cartesian, with a resolution of $N_x^3$,
which is generally $128^3$.
The Lagrangian triangular grid used to represent the interface is
constructed in such a way that the sides of the triangles remain
close to the length of the diagonals of the cubic Eulerian mesh.
The grid spacing $\Delta x=4R/N_x$ for the capillary cases should be
compared to the circumference $2\pi R$ of the drop
and to the approximate wavelengths $2\pi R/\ell$.
The radius and circumference are spanned by
$R/\Delta x = N_x/4$ and $2\pi R/\Delta x=N_x \pi/2$,
which are 32 and 201, respectively for the $128^3$ grid.
The number of gridpoints per wavelength $N_x\pi/(2\ell)$
goes from 201 for $\ell=1$ to 33 points for $\ell=6$ for this grid.
The interface is represented by approximately $4\pi R^2/(\Delta x)^2
= N_x^2\pi/4$ points, i.e. 12,868 points for $N_x=128$.
For the gravitational cases, the radius of the bounding sphere is 
$2.4\Rd$, 
leading to slightly different but similar numbers for the resolution,
e.g. $R/\Delta x=26.7$ rather than 32.
We have also carried out simulations with grids of $256^3$
(to confirm our results) and of $64^3$ (for visualisation of our patterns).
Parallelization is achieved through domain decomposition, 
in which each subdomain is assigned to its own process thread.
Here, we use $512$ subdomains or processes,
each with $16^3$ or $32^3$ gridpoints for the global resolutions
of $128^3$ and $256^3$, respectively.


The time step $\Delta t$ is chosen at each iteration in order to satisfy a criterion based on
\begin{equation}\label{eq:steps}
\left\{\Delta t_{\rm CFL}, \Delta t_{\rm int}, \Delta t_{\rm vis}, \Delta t_{\rm cap}\right\}
\end{equation}
which ensures stability of the calculations.  These bounds are defined by:
\begin{equation}\label{eq:Time_steps}
\begin{array}{ll}
\displaystyle{\Delta t_{\rm CFL} \equiv \min_j \left( \min_{\rm domain} \left(\frac{\Delta x_j}{u_j}\right) \right)} &
\displaystyle{\Delta t_{\rm int} \equiv \min_j \left( \min_{\Gamma(t)} \left( \frac{\Delta x_j}{\|{\bf V}\|}\right) \right)}\\
\displaystyle{\Delta t_{\rm vis} \equiv \min\left(\frac{\rho_{2}}{\mu_{2}},\frac{\rho_{1}}{\mu_{1}}\right)\frac{\Delta x_{\min}^2}{6}}&
\displaystyle{\Delta t_{\rm cap} \equiv \frac{1}{2}\bigg(\frac{(\rho_{1} + \rho_{2}) \Delta x_{\min}^3}{\pi \sigma} \bigg)^{1/2}}
\end{array}
\end{equation}
where $\Delta x_{\min} = \min_j \left(\Delta x_j \right)$.
In our simulations, this minimum is realized by
$\Delta t_{\rm CFL}\approx\Delta t_{\rm int}$ for our gravitational waves
and by $\Delta t_{\rm cap}\approx\Delta t_{\rm vis}$ for our capillary waves.

\subsection{Spherical harmonic transform}
\label{sec:sht}

In order to analyze the shape of the drop quantitatively,
we will compute and present the time-dependent spectral coefficients,
obtained from the decomposition
\begin{equation}
\zeta(\theta,\phi,t) = \sum_{\ell=0}^\infty \sum_{m=-\ell}^{m=\ell} \zeta_\ell^m(t)Y_\ell^m(\theta,\phi)
\label{eq:defint}\end{equation}
where $\zeta$ is the distance from the domain center. We compute both $|\zeta_\ell^m(t)|$  and
\begin{equation}
\zeta_\ell(t) \equiv \left[\sum_{m=0}^\ell |\zeta_\ell^m(t)|^2\right]^{1/2}
\label{eq:coeffylm}
\end{equation}
We note that, according to the normalization convention we use, 
\begin{equation}
  \zeta_0(t)=\zeta_0^0=\int_0^{2\pi} d\phi\int_0^{\pi} d\theta\: \sin\theta \:\zeta(\theta,\phi,t) \,Y_0^0(\theta,\phi)
\label{eq:zeta0}\end{equation}
i.e. $\zeta_0(t)$ is $\sqrt{4\pi}$ times the spherically averaged distance of the 
interface from the origin.

We calculate the spherical harmonic transform as follows.
The three-dimensional
Lagrangian interface consists of triangles composed of points
$\left(x_i, y_i, z_i \right)$ on the fixed Cartesian grid.  We transform each of
these points to spherical coordinates
$\left(\zeta_i, \theta_i, \phi_i\right)$ and 
interpolate the function $\zeta(\theta_i,\phi_i) = \zeta_i$
onto a regular spherical grid $(\theta_j,\phi_k)$,
discretizing the latitudinal interval
$[0,\pi]$ and longitudinal interval $[0,2\pi]$ with $N_\theta = 80$
and $N_\phi = 80$ points.  Using the discrete data
$\zeta(\theta_j,\phi_k)$ and truncating the series \eqref{eq:defint} at
four times the dominant value of $\ell$, we compute the spherical
harmonic coefficients $\zeta_\ell^m$ by the method of least squares
\citep{Politis,politis_phd}.  
This procedure is carried out for each sampled temporal
snapshot of the interface. We note that the spherical harmonic coefficients
are weighted averages of the surface height over the entire drop surface,
and hence more accurate than the individual surface height values.

The spectrum in $m$ (but not in $\ell$)
depends on the orientation. We can rotate a pattern by localizing
the coordinates $(\theta,\phi)$ of one of its features, 
such as a maximum, rotating in longitude $\phi$ to place it
in the $(x,z)$ plane and then rotating
in colatitude $\theta$ to place this feature at the north pole.
This allows us to interpret our spectra by using the 
explicit representations given by \cite{Busse1975} of 
various patterns in terms of $Y_\ell^m$.
Conversely, if the shape of the pattern is known or constant,
the spectrum in $m$ can be used to track changes in its orientation. 

\subsection{Validation}

\begin{figure}
\centerline{\includegraphics[width=0.8\textwidth,clip]{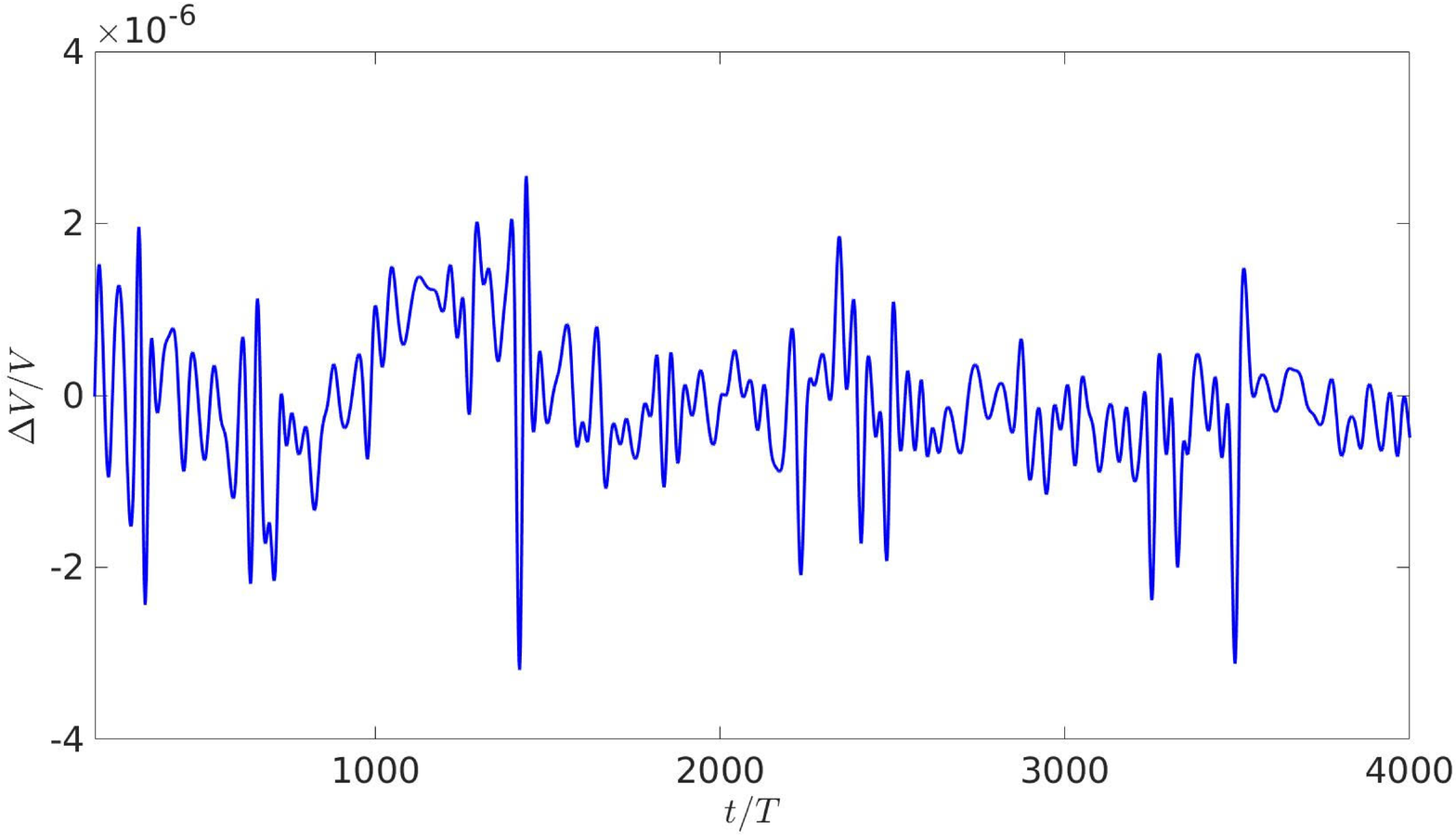}}
\caption{Time evolution of normalized deviation of
  $\Delta V (t)/V$ for a simulation with $\ell=6$.
  The value of $|\Delta V|/V$ remains less than about $3\times 10^{-6}$
  throughout the simulation.}
	\label{fig:volume_validation}
\end{figure}

The front-tracking approach inherently conserves mass to high accuracy
compared to other numerical interface methods \citep{ShJu09}.
We confirm mass conservation by showing the fractional deviation of
the volume.
Figure \ref{fig:volume_validation} shows the time evolution of the bounding envelope of
$\Delta V/V$
for a simulation with resolution $128^3$. The fluid parameters are those of section
\ref{sec:phys_par}
and table \ref{tab:forcing_response} for the $\ell=6$ case.
(See figure \ref{fig:S_ell2_L} for an illustration
of the extraction of the bounding envelope.)
We see that $|\Delta V(t)|/V$ remains less than about $3\times 10^{-6}$
throughout the simulation.

\begin{figure}
\centerline{\includegraphics[width=0.8\textwidth,clip]{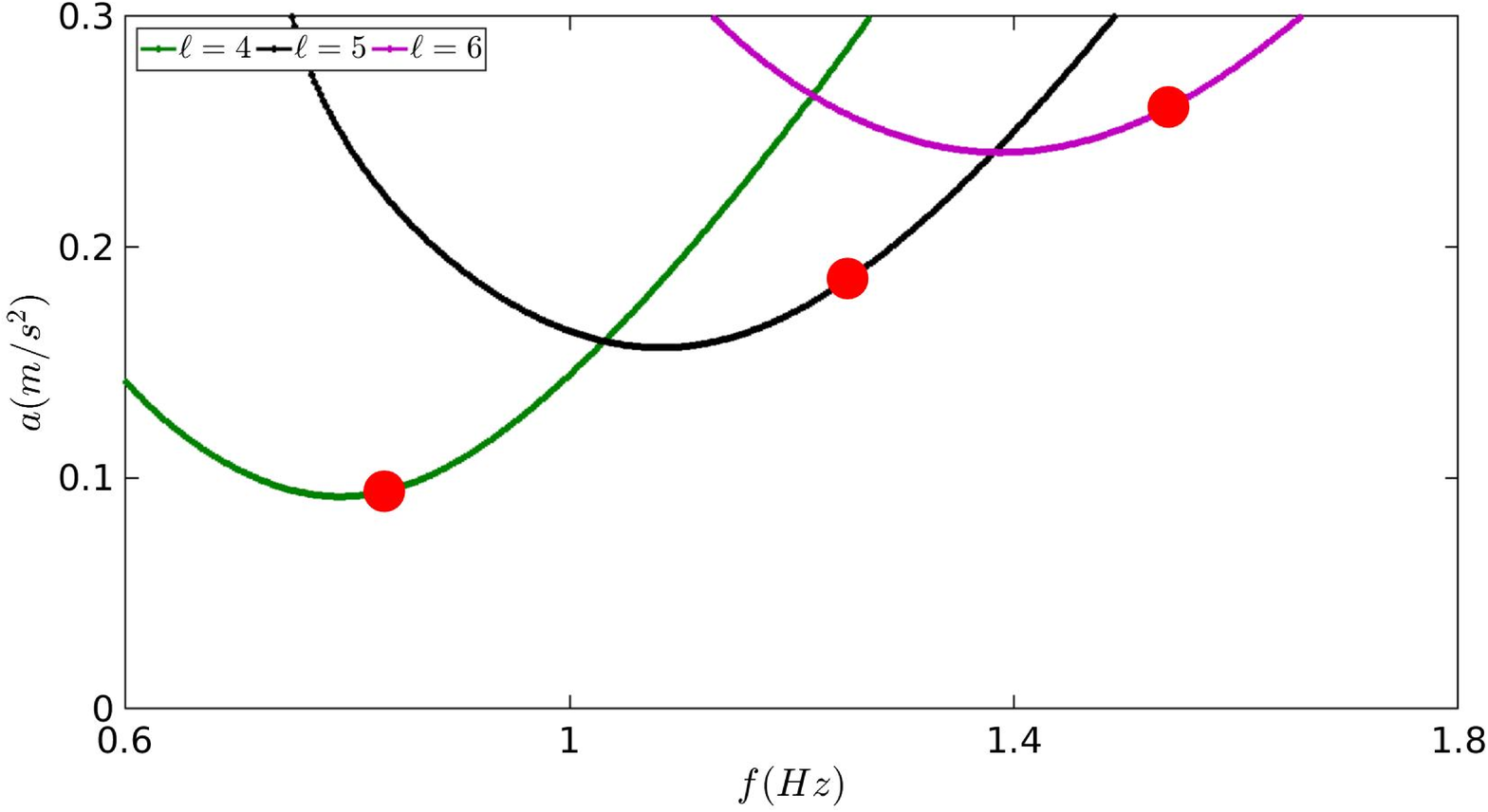}}
\caption{Comparison between theoretical and computed threshold $a_c$
  Curves show theoretical thresholds 
  calculated from Floquet theory \citep{Ali1} for $\ell=4$, 5, 6, 
  while crosses show results from numerical simulations with resolution $256^3$
  at the indicated frequencies calculated by the method shown in figure \ref{fig:valid_slopes_rates}.}
	\label{fig:validation}
\includegraphics[width=\textwidth]{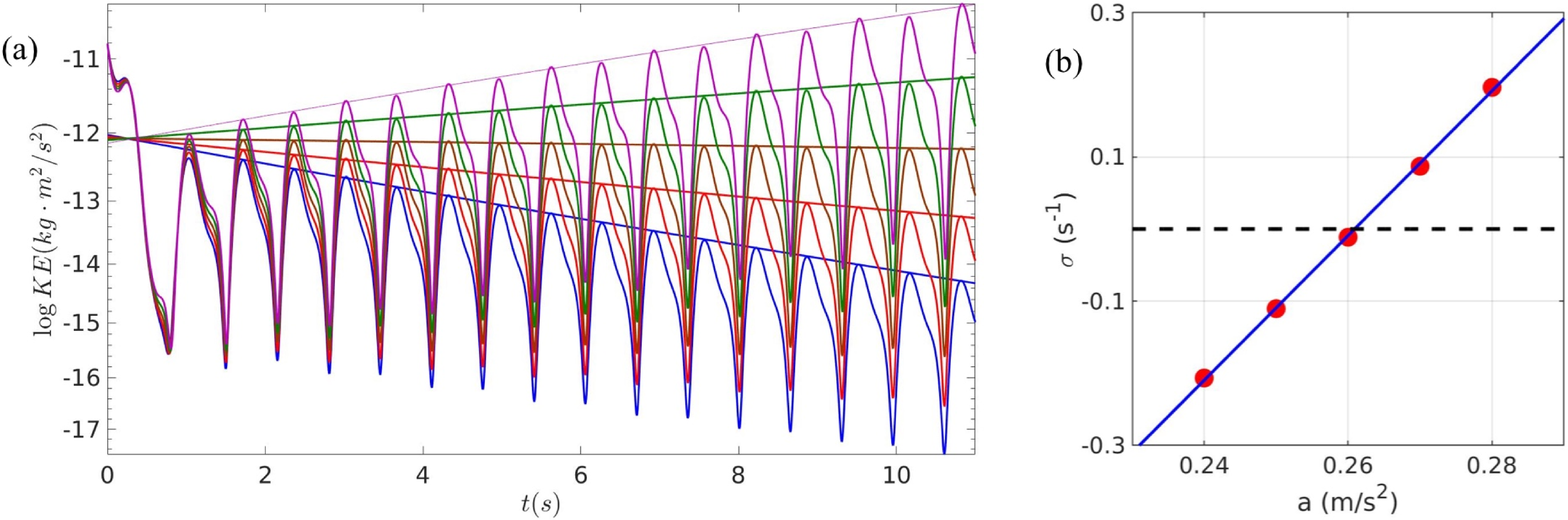}
\caption{Calculation of thresholds. a) Timeseries of interface height
  oscillate and grow or decay exponentially, depending
  on the imposed acceleration amplitude,
  as shown by the linear time dependence of the logarithms of the peaks.
  b) The slopes of the logarithms of the timeseries maxima constitute
  the growth or decay rates,
  which are interpolated to determine the acceleration amplitude threshold
  for $\ell=6$ shown as the rightmost cross in figure \ref{fig:validation}.}
	\label{fig:valid_slopes_rates}
\end{figure}
In our previous investigation \citep{Ali1}, we extended the method
of \cite{KT1994} for computing the Faraday threshold via Floquet
theory to a spherical geometry.  In figure \ref{fig:validation} and
table \ref{tab:validation} we compare these theoretical results with
thresholds obtained by interpolating growth rates from numerical
simulations, as shown in figure \ref{fig:valid_slopes_rates}.  The
parameters are as previously stated for capillary waves in section
\ref{sec:phys_par}, except that we increase the viscosity to
$\nu_d=10^{-4} m^2s^{-1}$.  For $\ell=6$, the error in the threshold is about
1\% for a resolution of $128^3$ and only about 0.1\% for $256^3$.

\begin{table}
\begin{center}
\begin{tabular}{ccccc}
\hline 
\begin{tabular}{c}Spherical mode $\ell$\\~\end{tabular} & \begin{tabular}{c}Resolution\\~\end{tabular} & \begin{tabular}{c}Theoretical\\ $m\,s^{-2}$\end{tabular} & \begin{tabular}{c}Numerical\\$m\,s^{-2}$\end{tabular} & $\dfrac{|\Delta a_c|}{a_c}$(\%) \\
\hline 
  6 & $128^3$ & 0.2604~ & 0.2640~ & 1.38 \\
  5 & $128^3$ & 0.1857~ & 0.1886~ & 1.52 \\
  4 & $128^3$ & 0.09415 & 0.0919~ & 2.39 \\
  \hline
6 & $256^3$ & 0.2604~ & 0.2607~ & 0.08 \\
5 & $256^3$ & 0.1857~ & 0.1864~ & 0.34 \\
4 & $256^3$ & 0.09415 & 0.09424 & 0.10 \\
\hline 
\end{tabular}
\end{center}
\caption{Comparison between theoretical and computed thresholds $a_c$ for various
  $\ell$ values and numerical resolutions.}
	\label{tab:validation}
\end{table}

\begin{figure}
\includegraphics[width=\textwidth,clip]{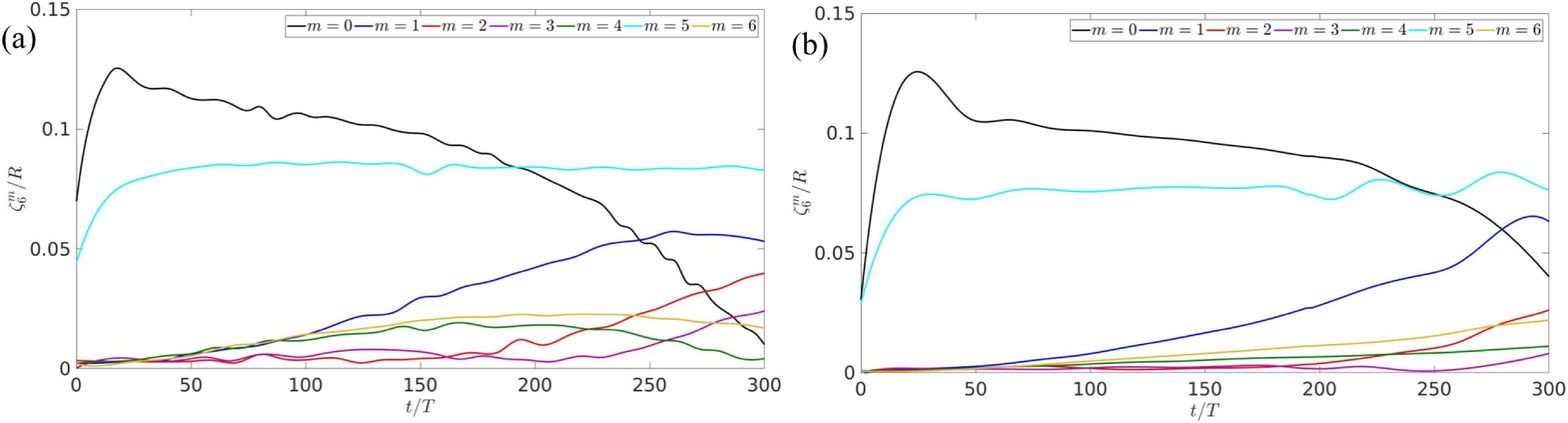}
\caption{Comparison of evolution of spherical harmonic spectra $Y_6^m$ for
  resolutions (a) $128^3$ and (b) $256^3$.}
	\label{fig:valid_mspectra}
\end{figure}

Figure \ref{fig:valid_mspectra} compares the time evolution 
of the envelope of the spherical harmonic coefficients
$|\zeta_\ell^m|$ for capillary wave simulations using resolutions
$128^3$ and $256^3$.  The fluid parameters are again those of section
\ref{sec:phys_par}
and table \ref{tab:forcing_response} for the $\ell=6$ case.
  Starting from an icosahedral initial condition, 
  the solution evolves similarly for the two spatial resolutions.
  The main difference is that near-zero modes grow more slowly
  for the higher resolution, which can be understood
  as a manifestation of the lower level of noise introduced
  in the simulation. 

\begin{figure}
\centering
\begin{tabular}{c}
\includegraphics[width=\textwidth]{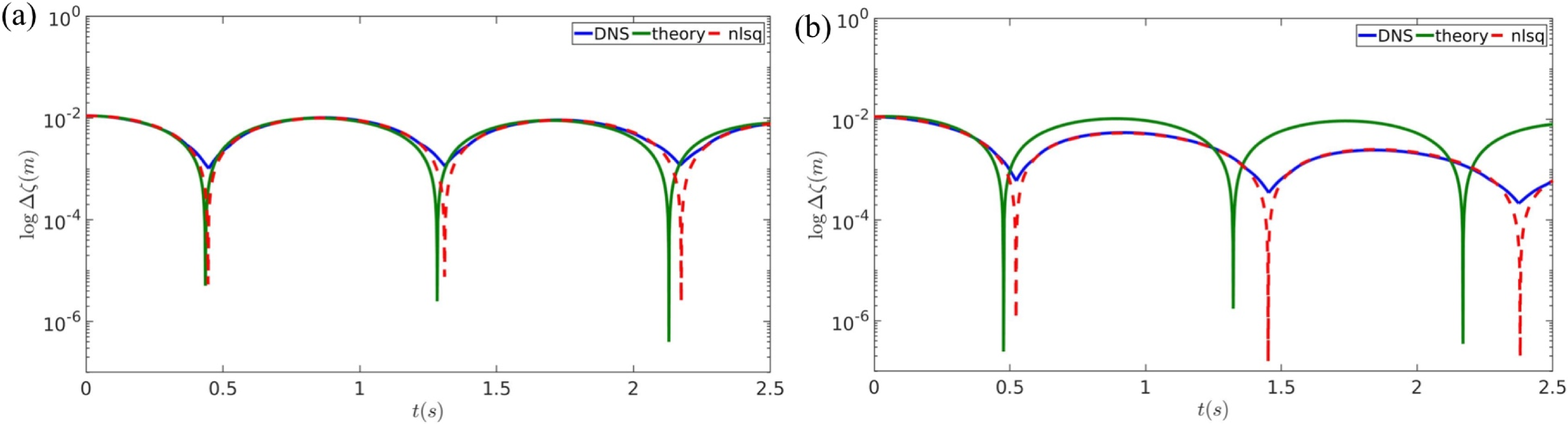}
\end{tabular}
\caption{Free decay of a shape-deformed drop to the spherical shape. The evolution of $\Delta \zeta \equiv \zeta_{max}-\zeta_{min}$ with time is shown for $\ell=5$ for a resolution of $256^3$. The blue curve corresponds to the output from numerical simulations which is fitted by a nonlinear least-square method (dashed red). The green curve corresponds to the small-viscosity-limit prediction for viscosity (a) $\nu_d=10^{-5} m^2s^{-1}$ and (b) $\nu_d=10^{-4} m^2s^{-1}$.}
    \label{fig:decL5a3}
\end{figure}

\begin{table}
\begin{center}
\begin{tabular}{cccccccccc}
\hline 
Viscosity  & Spherical Mode & Numerical & \multicolumn{2}{c}{Low-viscosity} & \multicolumn{2}{c}{Numerical} & \multicolumn{2}{c}{Deviation (\%)} \\
& & resolution & \multicolumn{2}{c}{theory} & & \\
  \hline
  $\begin{array}{c}{\nu_d}\\(m^2s^{-1})\end{array}$ & $\begin{array}{c}\ell\\~\end{array}$ & & $\begin{array}{c}\lambda\\(s^{-1})\end{array}$ & $\begin{array}{c}\omega^\prime\\(s^{-1})\end{array}$ & $\begin{array}{c}\lambda\\(s^{-1})\end{array}$ & $\begin{array}{c}\omega^\prime\\(s^{-1})\end{array}$
  & $\dfrac{|\Delta\lambda|}{\lambda}$ & $\dfrac{|\Delta\omega^\prime|}{\omega^\prime}$ \\
\hline
$10^{-5}$ & 5 & $128^3$ & 0.122 & 3.711 & 0.113 & 3.607 & 7.53 & 2.83 \\
$10^{-5}$ & 5 & $256^3$ &  &  & 0.114 & 3.635 & 6.71 & 2.06 \\
\hline
$10^{-4}$ & 5 & $128^3$ & 1.222 & 3.507 & 0.799 & 3.607 & 34.6 & 4.33 \\
  $10^{-4}$ & 5 & $256^3$ &  &  & 0.805 & 3.381 & 34.1 & 3.56 \\
  \hline
$10^{-5}$ & 4 & $128^3$ & 0.075 & 2.661 & 0.070 & 2.615 & 6.66 & 1.72 \\
$10^{-5}$ & 4 & $256^3$ &  &  & 0.073 & 2.632 & 3.34 & 1.09 \\
\hline
$10^{-4}$ & 4 & $128^3$ & 0.750 & 2.555 & 0.523 & 2.497 & 30.2 & 2.25 \\
$10^{-4}$ & 4 & $256^3$ &  &  & 0.529 & 2.512 & 29.5 & 1.68 \\
\hline
\end{tabular}
\end{center}
\caption{Comparison between decay rates and nonlinear frequencies
  from simulations and from low-viscosity theory of \eqref{eq:Lambform}-\eqref{eq:Lamb}.}
	\label{tab:Lamb}
\end{table}

We also validated our numerical code by testing the free decay of a
perturbed sphere in the unforced case (i.e. $a=0$, no imposed oscillatory forcing)
under the influence of surface tension.
In the absence of viscosity, \cite{Rayl1879} showed that 
a drop initialized with a perturbation of spherical
wavenumber $\ell$ oscillates with frequency $\omega$ where
\begin{equation}
\omega^2=\frac{\sigma}{\rho}\frac{\ell(\ell-1)(\ell+2)}{R^3}
\label{eq:Rayleigh}\end{equation}
For small kinematic viscosity $\nu_d$,
\cite{Lamb1932} showed that the oscillation amplitude decays to the
spherical rest state according to 
\begin{equation}
\zeta_\ell(t) \propto e^{-\lambda t } | \cos(\omega^\prime t) |
\label{eq:Lambform}\end{equation}
\noindent where the decay rate and the oscillation frequency are given by
\begin{equation}
\lambda = \frac{\nu_d (\ell-1)(2\ell+1)}{R^2} \hspace*{3cm}
\omega^\prime = \left( \omega^2 - \lambda^2 \right)^{1/2}.
\label{eq:Lamb}\end{equation}
Figure \ref{fig:decL5a3} shows the evolution of
$\Delta \zeta \equiv \zeta_{max}-\zeta_{min}$ obtained via numerical
simulations with a resolution of $256^3$ of capillary waves with the
fluid parameters of section \ref{sec:phys_par} initialized with
axisymmetric perturbations $Y_5^0$ for two different viscosities,
along with curves obtained from
the low-viscosity theory of \eqref{eq:Lambform}-\eqref{eq:Lamb}.
Table \ref{tab:Lamb} gives the values of
the decay rates and nonlinear frequencies from figure \ref{fig:decL5a3},
along with additional cases in which the spherical mode and the spatial resolution
have been varied.
As could be expected, the deviation is lowest when the resolution is higher ($256^3$)
so that the simulation is more accurate, and when the viscosity is lower ($\nu_d=10^{-5}ms^{-2}$)
so that the low-viscosity theory is more applicable.
This best-case deviation is about 7\% for $\lambda$ and 2\% for $\omega^\prime$
for $\ell=5$ and only about 3\% for $\lambda$ and 1\% for $\omega^\prime$ for $\ell=4$.
%

\subsection{Survey of cases studied}
\label{sec:cases}

Table \ref{tab:forcing_response} lists the simulations which we will describe in this paper.
As an initial condition, we perturb the spherical interface
by one of the following combinations of spherical harmonics:
\begin{subequations}
\begin{align}
&      \mbox{Axisymmetric} && \mbox{all } \ell && \zeta - \Rd \propto Y_\ell^0 \label{eq:axisym}\\
&\mbox{Tetrahedral}  && \ell=3&& \zeta-\Rd \propto Y_3^2 + {\rm c.c.}
  \label{eq:tetra}\\
&\mbox{Cubic} && \ell=4&& \zeta - \Rd \propto \sqrt{7}\: Y_4^0 + \sqrt{5}\:Y_4^4 + {\rm c.c.} \label{eq:cube}\\
&D_4 && \ell=5&& \zeta - \Rd \propto \sqrt{3}\: Y_5^0 + \sqrt{5} \: Y_5^4 + {\rm c.c.}\label{eq:D4}\\
&\mbox{Icosahedral} && \ell=6 &&\zeta - \Rd \propto \sqrt{11} \: Y_6^0  + \sqrt{14} \: Y_6^5 + {\rm c.c.} \label{eq:icosa}
\end{align}
\label{eq:formulas}\end{subequations}

\begin{table}
\begin{center}
\begin{tabular}{lll rr lll}
\hline 
& Temporal & & \multicolumn{2}{c}{Forcing} && Initial & Final\\
$\ell$ & Response & & Amplitude & Frequency && Condition & Pattern\\
\hline
\multicolumn{2}{c}{Gravitational}  & & $a/a_g$ & $\omega/\omega_g$ &\\
\hline
1 & harmonic & &  0.30 & 1.00 && spherical & translational\\
2 & harmonic & & 0.50 & 1.37 && axisymmetric& prolate-oblate\\
\hline 
\multicolumn{2}{c}{Capillary} & & $a/a_c$ & $\omega/\omega_c$ \\
\hline 
3 & subharmonic & & 13.53 & 13.21 && axisymmetric& tetrahedral\\
4 & subharmonic & & 5.07 & 16.61 && axisymmetric& axisymmetric\\
4 & subharmonic & & 5.07 & 16.61 && cubic & cubic-octahedral\\
5 & subharmonic & & 22.00 & 25.02 && axisymmetric& $D_4$\\
6 & subharmonic & & 16.07 & 30.63 && icosahedral & no stable pattern\\
\hline 
\end{tabular}
\end{center}
\caption{Parameters of imposed forcing and of the observed response.}
	\label{tab:forcing_response}
\end{table}

\begin{figure}
\centering
\includegraphics[width=1.1\textwidth,clip]{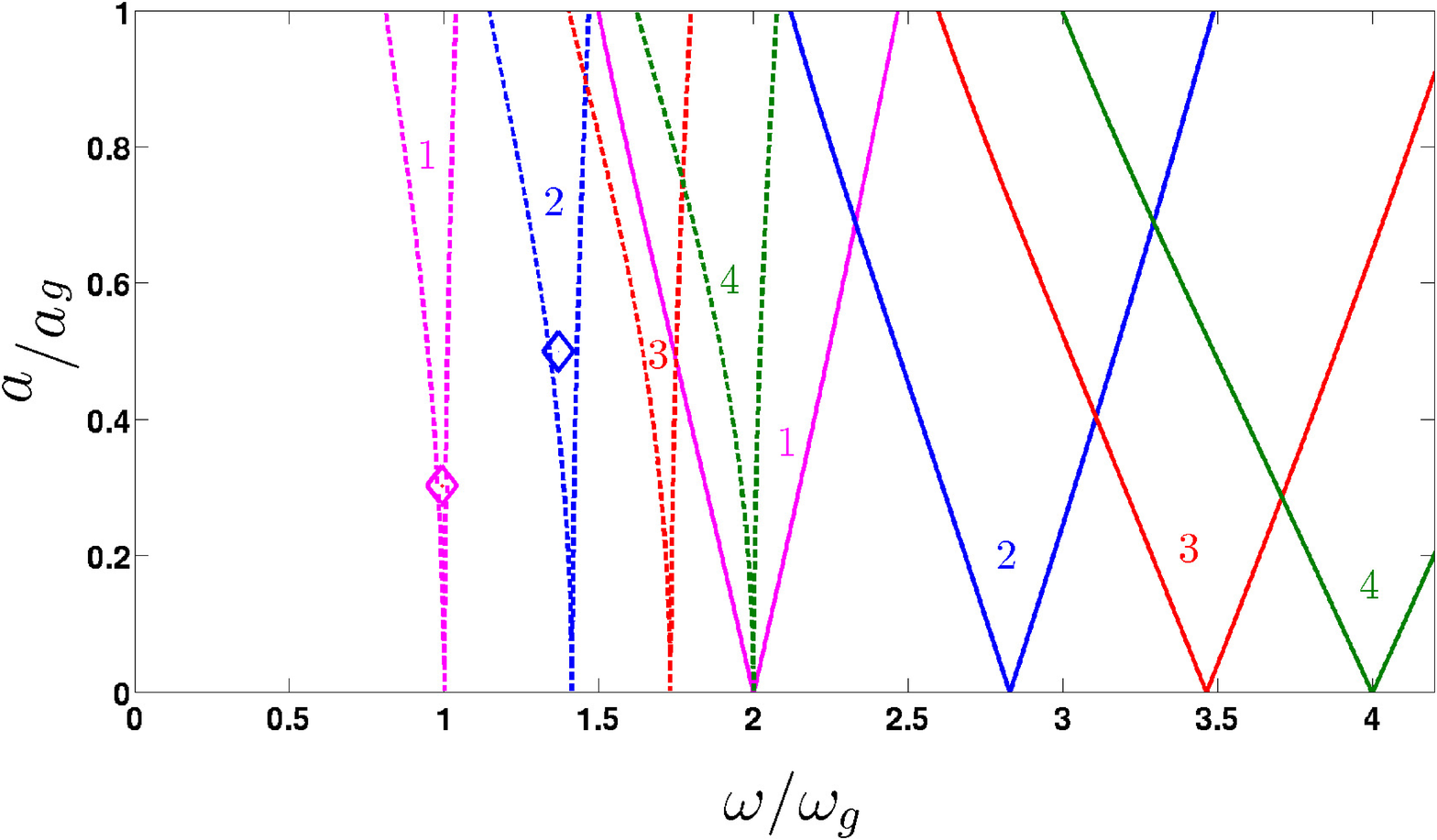} \\
\includegraphics[width=1.1\textwidth,clip]{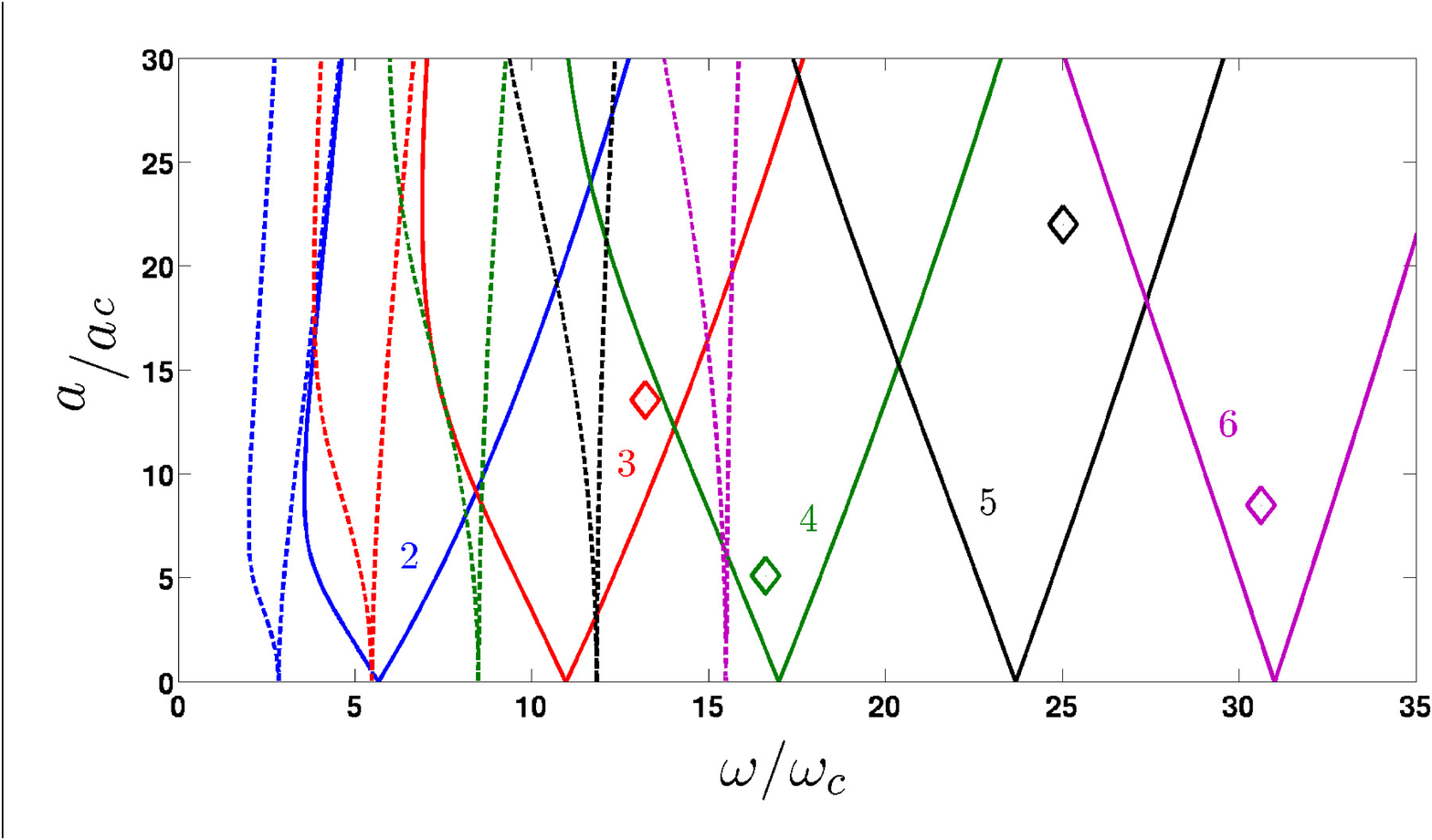}\\[-15cm]
\hspace{-1.\textwidth}
\begin{tabular}{l}
(a) \\[8cm] (b)\\[7cm]
\end{tabular}
\caption{Instability tongues 
  resulting from oscillatory forcing with amplitude $a$ and angular
  frequency $\omega$ for an inviscid drop with parameters given in
  table \ref{tab:forcing_response}. Diamond shapes designate the parameter values of
  our simulations.  Solid curves bound subharmonic tongues and dashed
  curves bound harmonic tongues.  (a)
  Tongues corresponding to gravitational instability with spherical
  wavenumbers $\ell=1$, 2, 3, 4 originate at $\omega/\omega_g =
  2\sqrt{\ell}/n$, with $n=1$ for subharmonic tongues and $n=2$
  for harmonic tongues.
(b) Tongues corresponding to
  capillary instability with spherical wavenumbers $\ell=2$, 3, 4, 5, 6
  originate at $\omega/\omega_c =2\sqrt{\ell(\ell-1)(\ell+2)}/n$.  
} \label{fig:BLUE_tongues}
\end{figure}

These formulas for functions with a given symmetry 
are given by \cite{Busse1975} and \cite{Riahi1984} for patterns
with a single $\ell$, aligned along the $z$ axis.
When the patterns are rotated to a different orientation,
the value of $\ell$ is conserved, but the combinations of $m$ values
change.
Patterns with other symmetries are also possible,
but are not used or not achieved here.
For $\ell=1$, the initial condition for the interface
is a sphere perturbed only by its representation on a triangular mesh.
In all cases, the initial velocity is zero.
The quantitative results we will present in the next section use a
resolution of $128^3$, confirmed by simulations with $256^3$.
Grids with $64^3$ are used to plot the visualisations of the drop.

The physical parameters are as given in section \ref{sec:phys_par}.
We study either gravity or capillary waves,
i.e.~a Bond number $\rho g R^2/\sigma$ of either infinity or zero. 
Figure \ref{fig:BLUE_tongues} locates the parameter values that we have
used for our simulations within the instability tongues 
for the gravitational or capillary cases, where 
\begin{subequations}\begin{eqnarray}
\omega_g^2 \equiv \frac{g}{R} && \qquad\omega_c^2 \equiv \frac{\sigma}{\rho R^3}\\
a_g \equiv R\omega_g^2 = g && \qquad a_c \equiv R\omega_c^2 =\frac{\sigma}{\rho R^2}
\end{eqnarray}\label{eq:nondim}\end{subequations}
See \cite{Ali1} for more details on this non-dimensionalization.
Because our viscosity is low, the tongues are very close to 
the inviscid ones; we plot the inviscid tongues for simplicity.

Frequencies were chosen to induce instabilities from $\ell = 1$ to $\ell = 6$,  as predicted from linear Floquet theory. 
In each case, the value of $\ell$ from the full three-dimensional nonlinear
numerical simulations agreed with the theoretical value.

\clearpage
\section{Results}

\subsection{Case $\lonset = 1$}
\label{sec:l=1}

\begin{figure}
\centering
\includegraphics[width=\textwidth,clip]{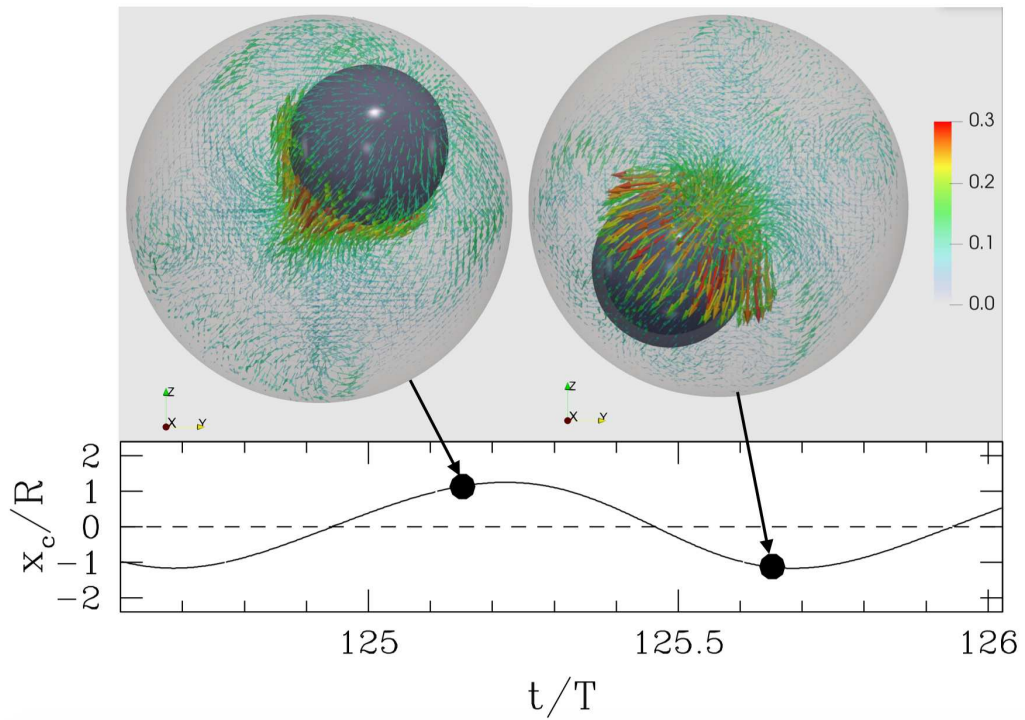}
\caption{Above: visualisation of $\lonset=1$ mode for spherical drop.  The drop is
  displaced alternately to the left and the right. Length and colors of
  arrows indicate the velocity magnitude (in m/s) of the surrounding air.
  Below: center of the drop as a function of time. Dots indicate
  the instants at which the visualisations are drawn.
See also supplementary movie 1.}
    \label{fig:l=1}
\end{figure}

We begin by presenting the $\lonset=1$ case in the purely
gravitational regime, i.e.~in the presence of a constant radial force
$g{\bf e}_r$ included in the time-periodic force \eqref{eq:grav} and
without surface tension.  Our simulations of this case exhibit a
periodic subharmonic translational motion of the sphere about its
original position, as shown in figure \ref{fig:l=1} and supplementary movie 1.
This explains why the $\lonset = 1$ case is prohibited for capillary waves:
translational motion does not involve deformation of the interface,
and so surface tension cannot act as a restoring force. (Note the
factor of ($\ell-1$) in formulas \eqref{eq:Rayleigh} and
\eqref{eq:Lamb}.) Although the drop arrives quite close to the
bounding sphere in this simulation, it does not touch it.

As stated by \cite{Busse1975}, there is no competition between
different patterns in this case.  All solutions are obtained by
rotation of a single axisymmetric pattern, which is expected to be
stable.  (The three spherical harmonics $Y_1^1,Y_1^0,Y_1^{-1}$ are
also related by rotation.)

\clearpage

\subsection{Case $\lonset = 2$}

\begin{figure}
\includegraphics[width=\columnwidth]{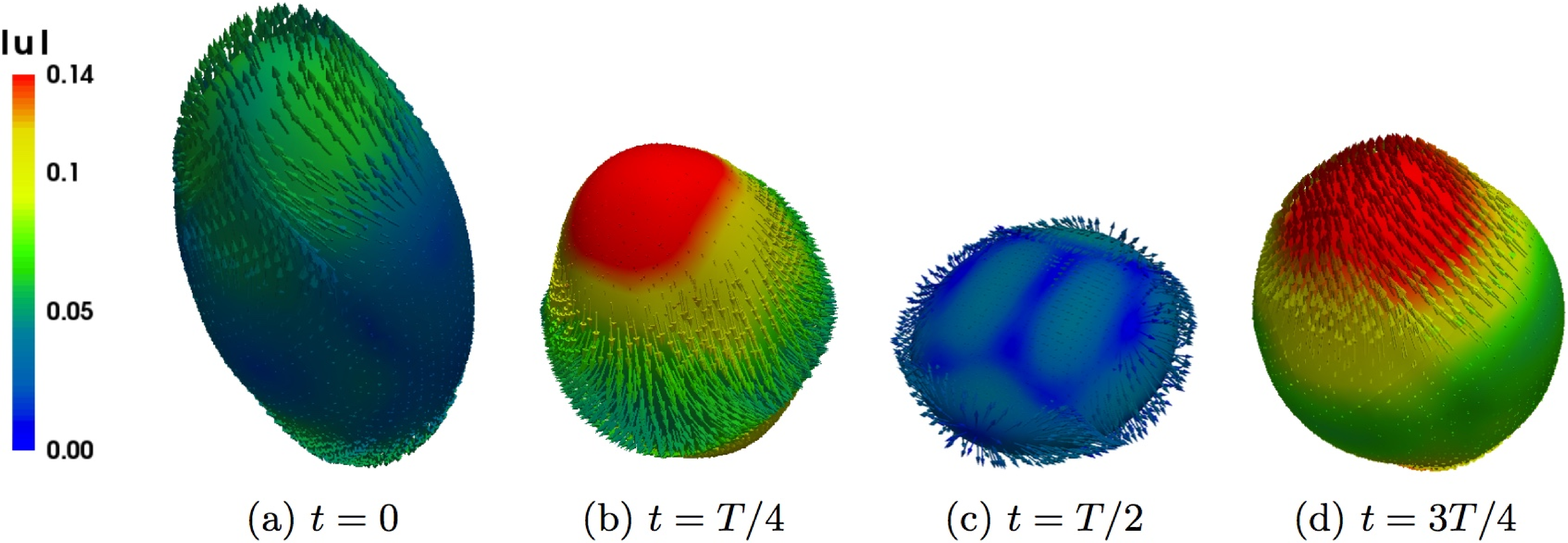}
\caption{Drop interface and velocity field for $\lonset=2$ prolate-oblate pattern of
  gravitational harmonic waves over one response period $T$. 
  During the prolate phase, the velocity is directed
  in the polar direction, while during the oblate phase it is directed
  in the equatorial direction.  Colors indicate the magnitude of the
  velocity (in m/s), which is maximal when the surface is least deformed and
  minimal where it is most deformed.  Only outward-pointing velocity
  vectors are shown; those pointing inwards are hidden by the opaque
  surface of the drop.}
    \label{fig:l=2}
%
\vspace*{0.5cm}
\includegraphics[width=\columnwidth]{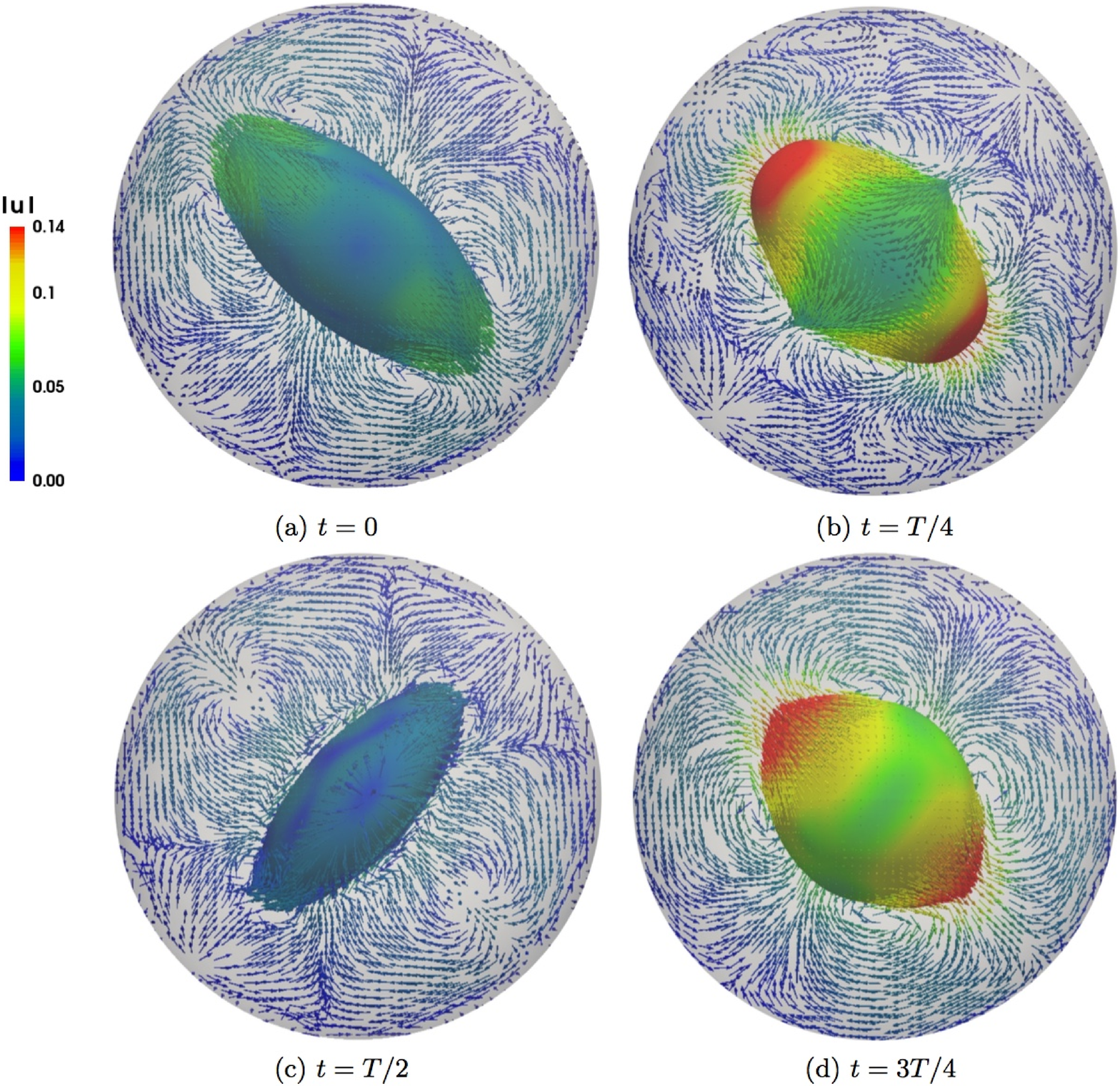}
\caption{Same as figure \ref{fig:l=2} but from a slightly different
  perspective and showing the outer bounding sphere and the
  velocity field in the outer fluid. See also supplementary movie 2.}
    \label{fig:l=2_out}
\end{figure}
\begin{figure}
  \centering
  \includegraphics[width=0.8\textwidth]{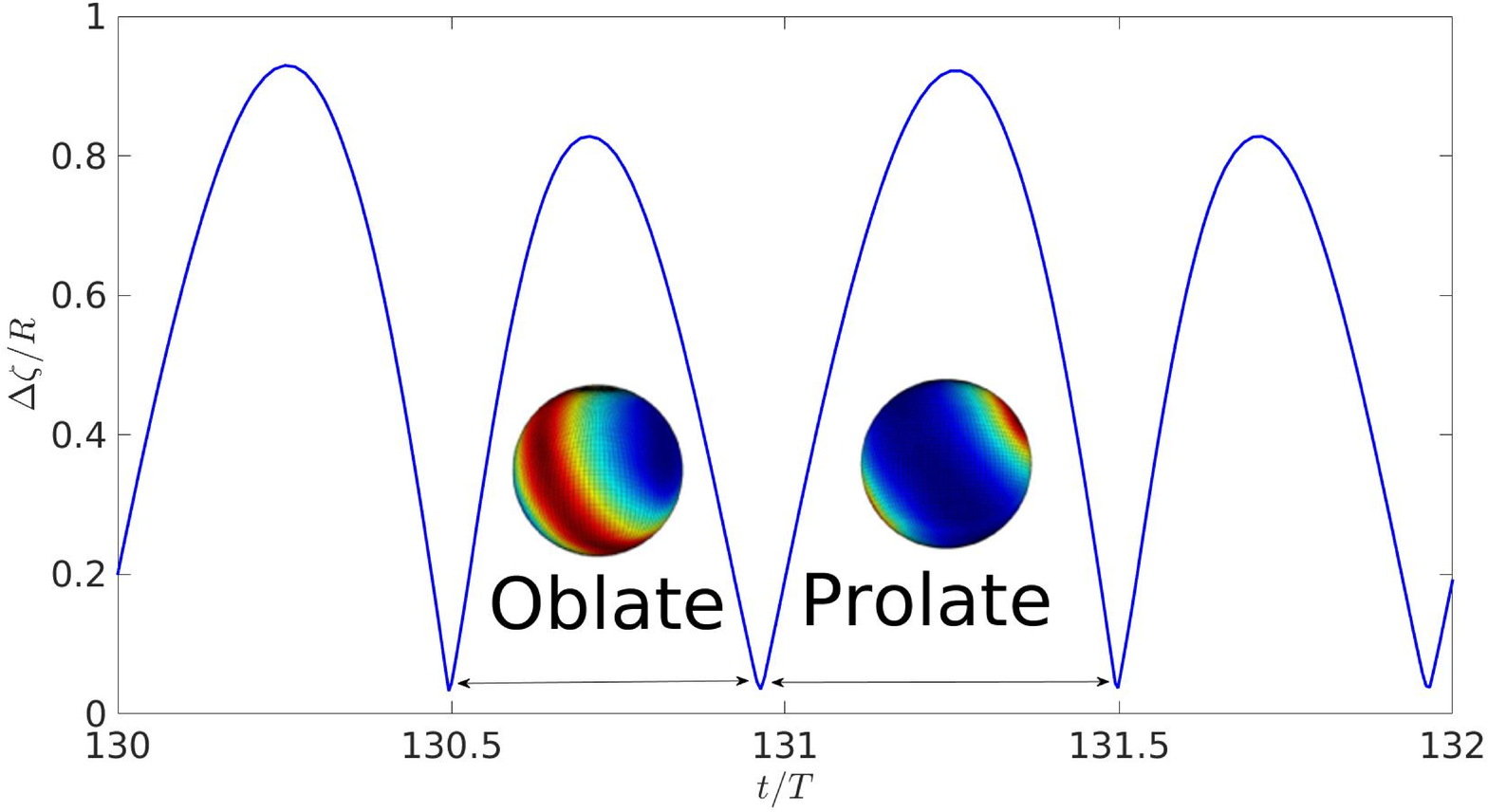}
  \caption{Timeseries of interface amplitude $|\zeta_{\rm max}-\zeta_{\rm min}|$
    for $\ell=2$.
Insets show the projection of the height $\zeta(\theta,\phi)$ on the sphere.
Prolate and oblate configurations have 
higher and lower maximum values of $|\zeta_{\rm max} - \zeta_{\rm min}|$,
respectively. The drop spends about
53\% of each period in the prolate configuration.}
  \label{fig:prol-obl}
%
  \includegraphics[width=0.8\textwidth]{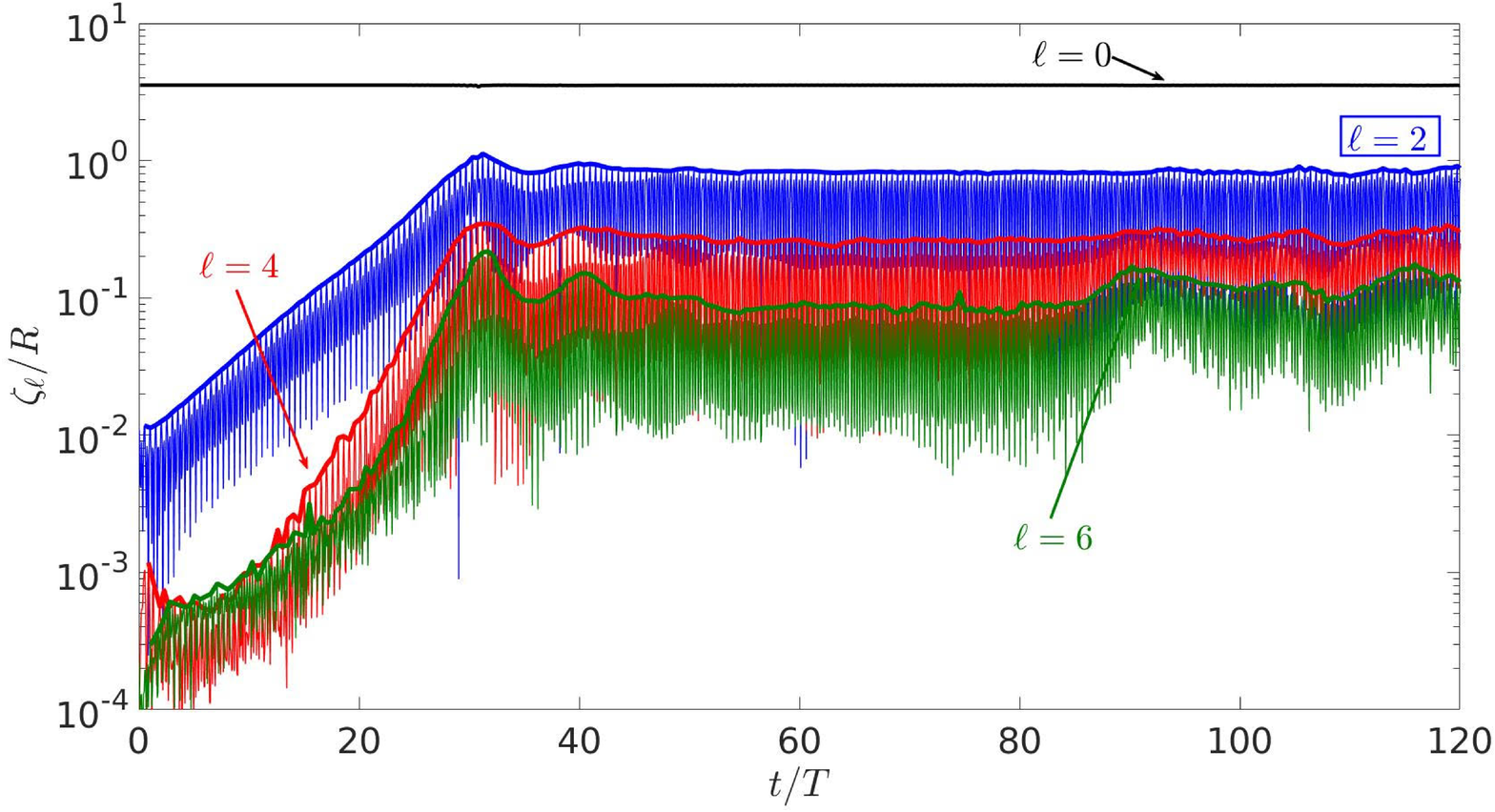}
  \caption{Timeseries of $\ell$ components for a pattern whose
    dominant mode is $\lonset=2$ (indicated by the boxed label).
    All multiples of $\lonset=2$ are present, as well as $\ell=0$,
    which is the constant average radius.
    Long-time evolution is visualized by envelopes (bold curves) of the
    rapidly oscillating timeseries.
    The growth rate of component $\zeta_4$ is about twice that of $\zeta_2$.}
\label{fig:S_ell2_L}
%
  \includegraphics[width=0.8\textwidth]{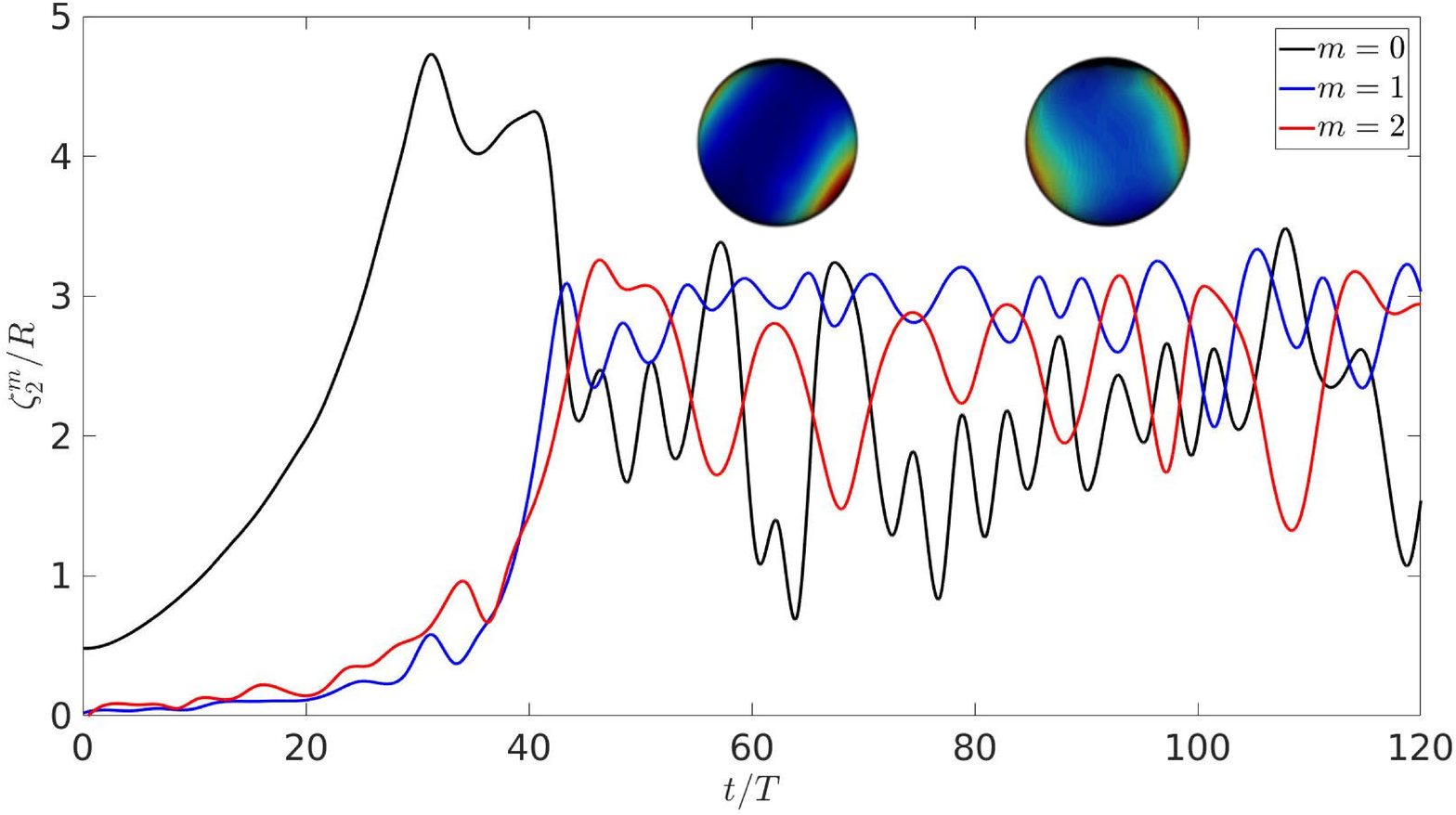}
  \caption{Timeseries of $m$ components for an $\ell=2$ pattern.
Initially, the drop is axisymmetric and the only component is $m=0$.
The drop remains axisymmetric, but its axis of symmetry 
tilts away from the $z$ axis and continues 
to rotate and oscillate, as manifested by alternating dominance of
$m=0$, 1, and 2.}
\label{fig:S_emm_L2}
\end{figure}

To produce a case in which $\lonset= 2$ is the dominant mode,
we have simulated harmonic oscillations in the gravitational regime. 
A sequence of figures from our numerical simulations showing the shape
and velocity field at the interface is shown in figure \ref{fig:l=2} and
figure \ref{fig:l=2_out}, which also shows the velocity field in the outer
fluid; see supplementary movie 2.  The interface remains axisymmetric, with two
principal axes of the same length, and a third of a different length.
Figure \ref{fig:l=2}(a) shows a prolate shape, like a rugby ball 
(the third axis is longer than the other two), while the shape in
figure \ref{fig:l=2}(c) is oblate, like a disk (the third axis is
shorter).  During the prolate phase, the velocity is directed in the
polar direction, while during the oblate phase it is directed in the
equatorial direction.  At maximum deformation, the velocity changes
direction, so that the magnitude of the velocity is lowest when the surface is most
deformed and highest when it is least deformed.  Because the interface
is opaque, velocity vectors directed inwards cannot be seen.

Figure \ref{fig:prol-obl} shows the amplitude
$|\zeta_{\rm max} - \zeta_{\rm min}|$ as a function of time.  The
prolate and oblate configurations are represented by higher and lower
maximum values of $|\zeta_{\rm max} - \zeta_{\rm min}|$, respectively.
Oblate-to-prolate oscillations have been studied extensively
\citep{TW1982,TB1983,Patz1991}. These authors observe that the drop
spends a longer time in the prolate than in the oblate configuration,
which agrees with our observation that the drop spends about 53\% of
each period in the prolate configuration.  This is explained by the
combination of two facts: the restoring force is governed by the
pressure at the curved surface, and the poles cover a smaller surface
than the equator.  The prolate rugby-ball-like form, with high
curvature at the poles, is therefore subjected to a smaller restoring
force than the oblate disk-like form, with high curvature at the
equator.  Hence it takes more time for the drop to return from the
prolate to the spherical shape than it takes to return from the oblate
shape.

In figures \ref{fig:S_ell2_L} and \ref{fig:S_emm_L2},
we present the spherical harmonic coefficients of the 
interface height $\zeta(\theta,\phi,t)$, calculated from 
equations \eqref{eq:defint} and \eqref{eq:coeffylm} as explained in section
\ref{sec:sht}.
Figure \ref{fig:S_ell2_L} shows the time evolution of 
$\zeta_\ell$, defined in \eqref{eq:coeffylm}.  Its most visible
feature is its rapid oscillation. 
In order to examine the dynamics over timescales much larger than
the forcing period $T$, we extract the
envelope of each $\zeta_\ell$, as shown by the bold curves in figure
\ref{fig:S_ell2_L}.  The variation of $\zeta_0$ cannot be seen on this scale,
since it is $\sqrt{4\pi}$ times the mean radius, which 
should be nearly constant due to incompressibility.
Although $\lonset=2$ is the dominant non-zero spherical wavenumber for
this case, other even $\ell$ (multiples of 2) are also present, 
generated by nonlinear interactions.
Standard bifurcation theory predicts that the growth
rate of the harmonic $n\ell$ should be approximately $n$ times that of
$\ell$, shown here by the fact that $\zeta_4$ grows about
twice as quickly as $\zeta_2$.

Although we began our simulations by perturbing the sphere with
an axisymmetric initial condition, proportional to $Y_2^0$,
the drop orientation quickly tilts away from
the $z$ axis, acquiring components with $m\neq 0$
while remaining axisymmetric about its own axis. 
The symmetry axis continues to rotate and to oscillate periodically,
manifested by the behavior of components $m=0$, 1 and 2, 
shown in figure \ref{fig:S_emm_L2}.
%
%
We have also obtained prolate-oblate oscillations in simulations (not
presented here) of subharmonic capillary waves with the same axisymmetric
initial condition.

\cite{Busse1975} and \cite{Chossat1991} show that the only allowed solution for
$\lonset=2$ is axisymmetric and unstable at onset. 
In our simulations, which are far from onset, 
we observe an axisymmetric pattern whose 
orientation continually changes.



\clearpage
\subsection{Case $\lonset = 3$}
\begin{figure}
\includegraphics[width=\columnwidth]{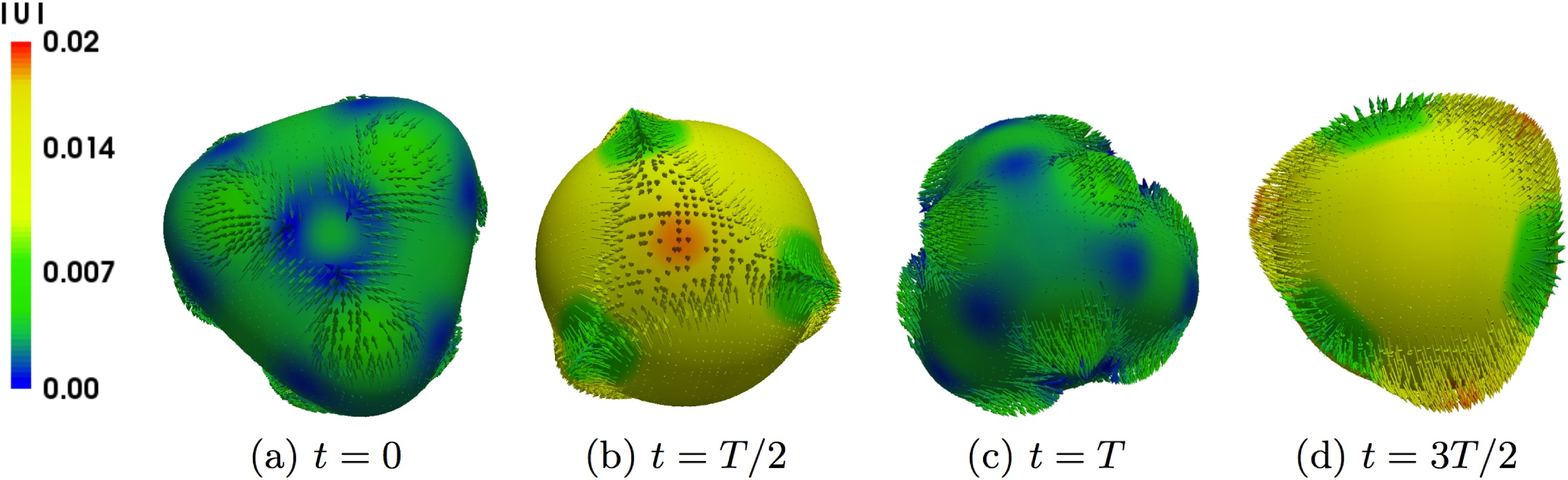}
\caption{Drop interface and corresponding velocity field for $\lonset=3$
  tetrahedral pattern seen in subharmonic capillary waves over one reponse period
  $2T$.  Colors indicate the magnitude of the velocity (in m/s), which is
  maximal when the surface is least deformed and minimal where it is
  most deformed.  Only outward-pointing velocity vectors are shown;
  those pointing inwards are hidden by the opaque surface of the
  drop. See also supplementary movie 3.}
    \label{fig:l=3}
%
\centering
\includegraphics[width=0.9\textwidth]{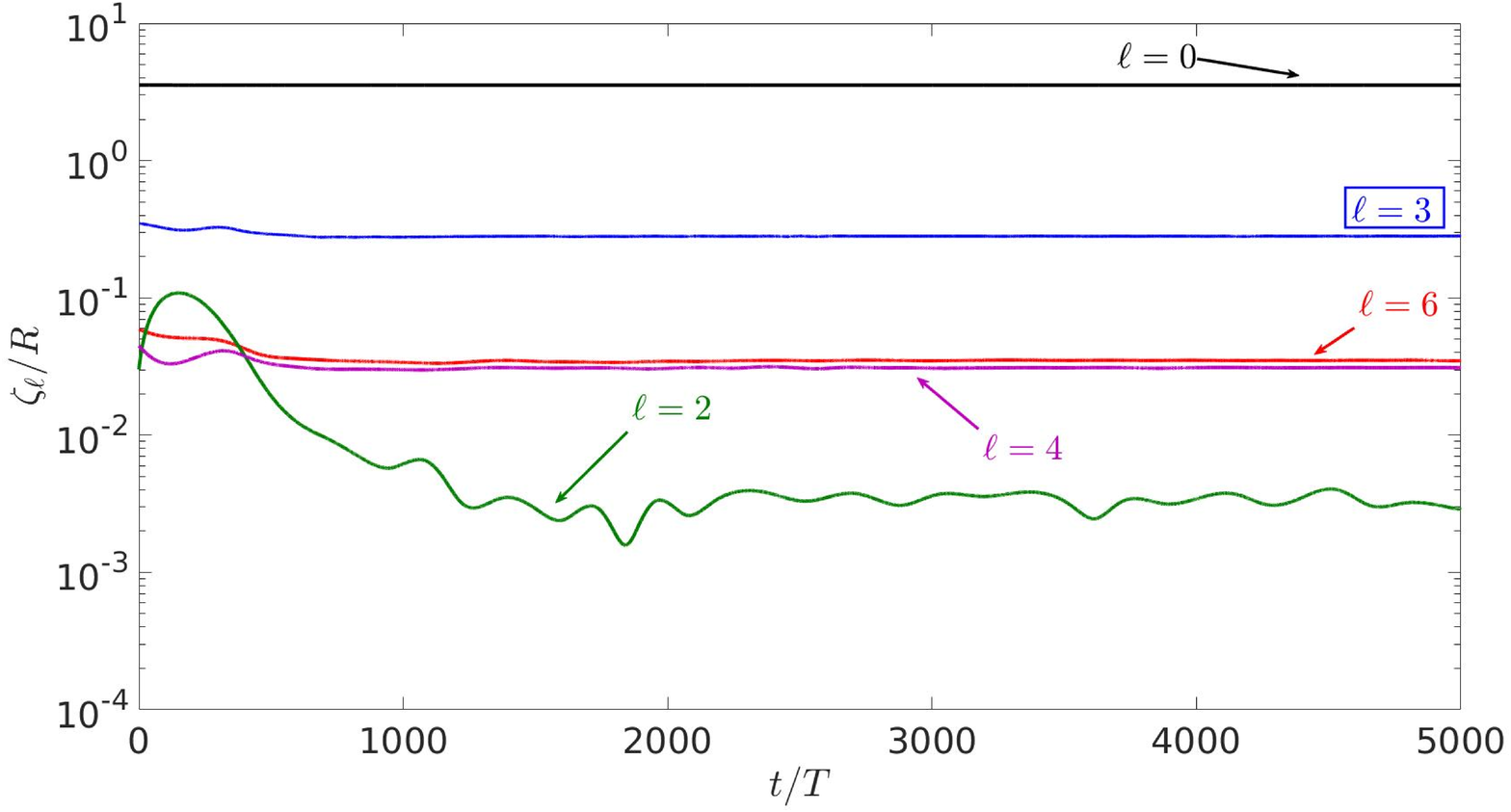}
\caption{Timeseries of $\ell$ components for a pattern whose
    dominant mode is $\lonset=3$.
The amplitude of mode $\ell=4$ is very close to that of the second
  harmonic $\ell=6$.}
    \label{fig:S_L3}
%
\centering
\includegraphics[width=0.9\textwidth]{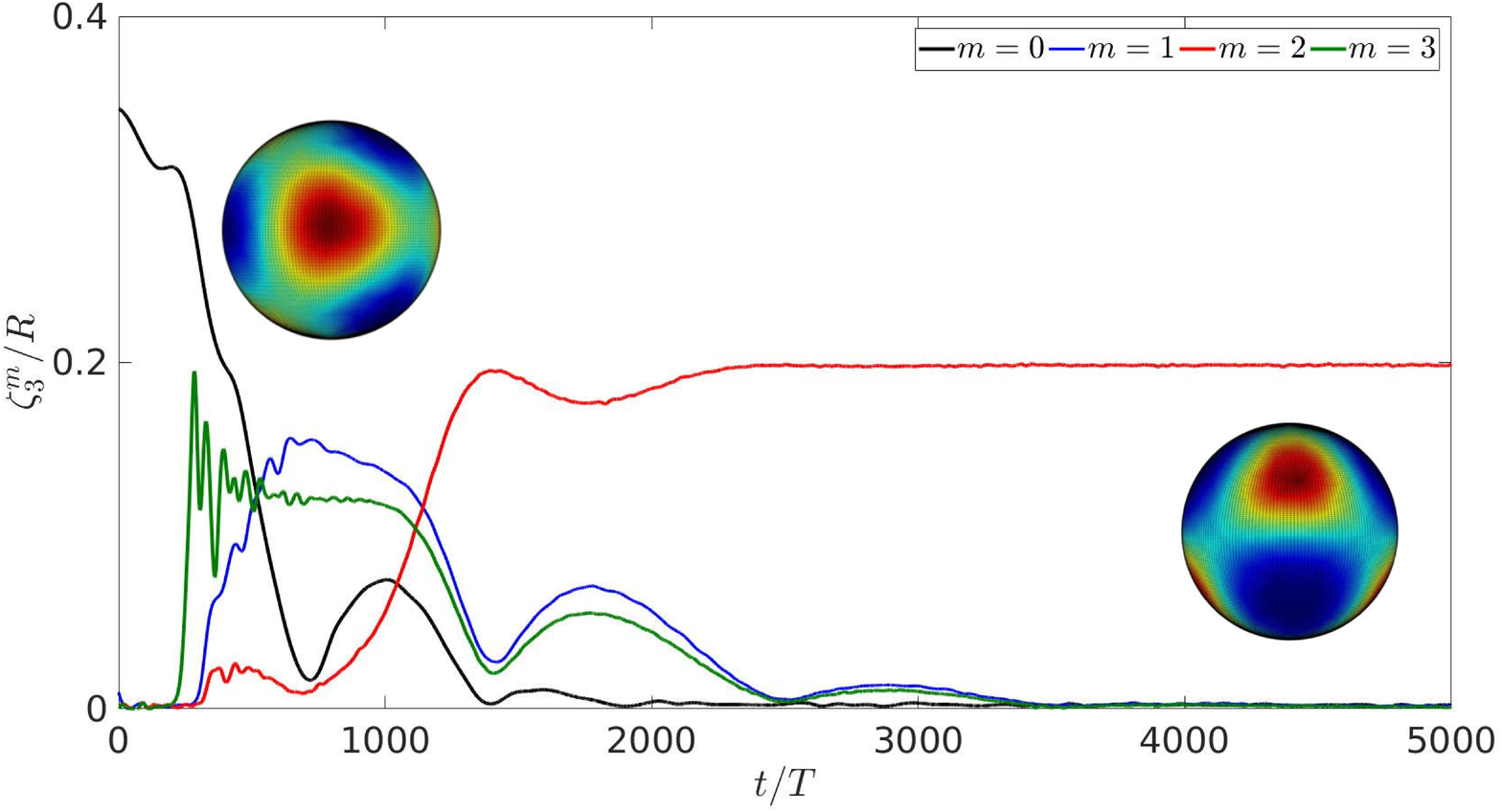}
\caption{Timeseries of $m$ components for an $\ell=3$ pattern.
  Initially the drop is axisymmetric. By $t/T\approx450$, the $m=3$
  component has risen and the shape is tetrahedral and tilted with
  respect to the $z$ axis. As the $m=2$ component increases, the
  pattern aligns with the $z$ axis.}
    \label{fig:S_emm_L3}
\end{figure}

For $\lonset=3$, there exist solutions with three possible symmetries:
axisymmetric, $D_6$, and tetrahedral. Either the $D_6$ or the tetrahedral
solution can be stable at onset \citep{Busse1975,Chossat1991}.
Starting from an axisymmetric initial condition, our solution
for this case rapidly develops tetrahedral symmetry,
as can be seen in figure \ref{fig:l=3} and supplementary movie 3,
which show subharmonic capillary oscillations. 
Over a half oscillation period,
the tetrahedron in figure \ref{fig:l=3}(a) reverses its orientation, 
as shown in figure \ref{fig:l=3}(c), since this polyhedron is self-dual.
The initial condition is axisymmetric, but the same behavior
is seen starting from a sphere slightly perturbed by numerical noise.

The pattern contains components other than $\lonset=3$ and its harmonics,
as illustrated in figure \ref{fig:S_L3}. In particular, an 
important $\ell=4$ component is present, of approximately
the same magnitude as the first harmonic $\ell=6$.
Figure \ref{fig:S_emm_L3} shows the rapid progression
from the axisymmetric initial condition
to a tetrahedral one, as component $m=0$ falls and the other components rise.
There follows a long phase during which the orientation of the tetrahedal
pattern varies. Eventually, by $t/T\approx 2000$, the 
tetrahedron aligns with the $z$ axis, with only $m=2$ remaining;
recall from equation \eqref{eq:tetra} that $Y_3^2 +{\rm c.c.}$
corresponds to a tetrahedron aligned with the $z$-axis \citep{Busse1975}.

\subsection{Case $\lonset = 4$}

\begin{figure}
\includegraphics[width=\textwidth]{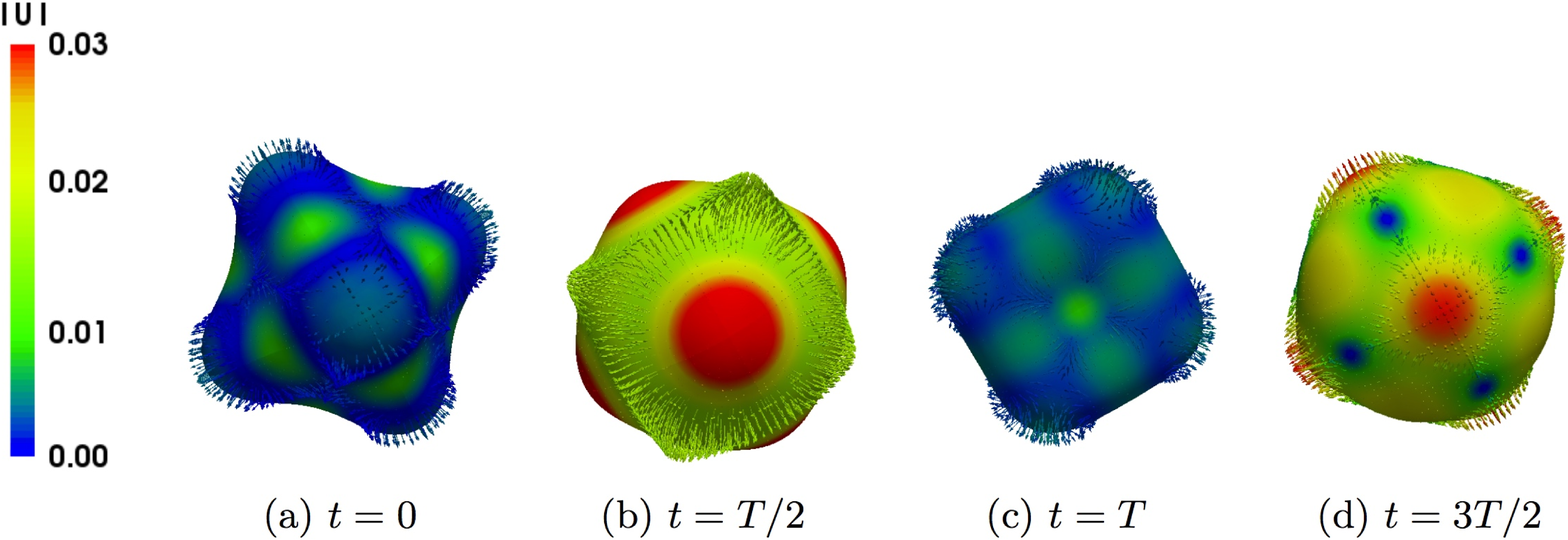}
\caption{Drop interface and velocity field for $\lonset=4$ cubic pattern seen
  in subharmonic capillary waves over one reponse period $2T$.  The
  interface oscillates between (a) an octahedron, with six maxima, and
  (c) a cube, with eight maxima.  Colors indicate the magnitude of the
  velocity (in m/s), which is maximal (minimal) when the surface is least
  (most) deformed.  Only outward-pointing velocity vectors are shown.
See also supplementary movie 4a.}
    \label{fig:l=4}
\includegraphics[width=0.9\textwidth]{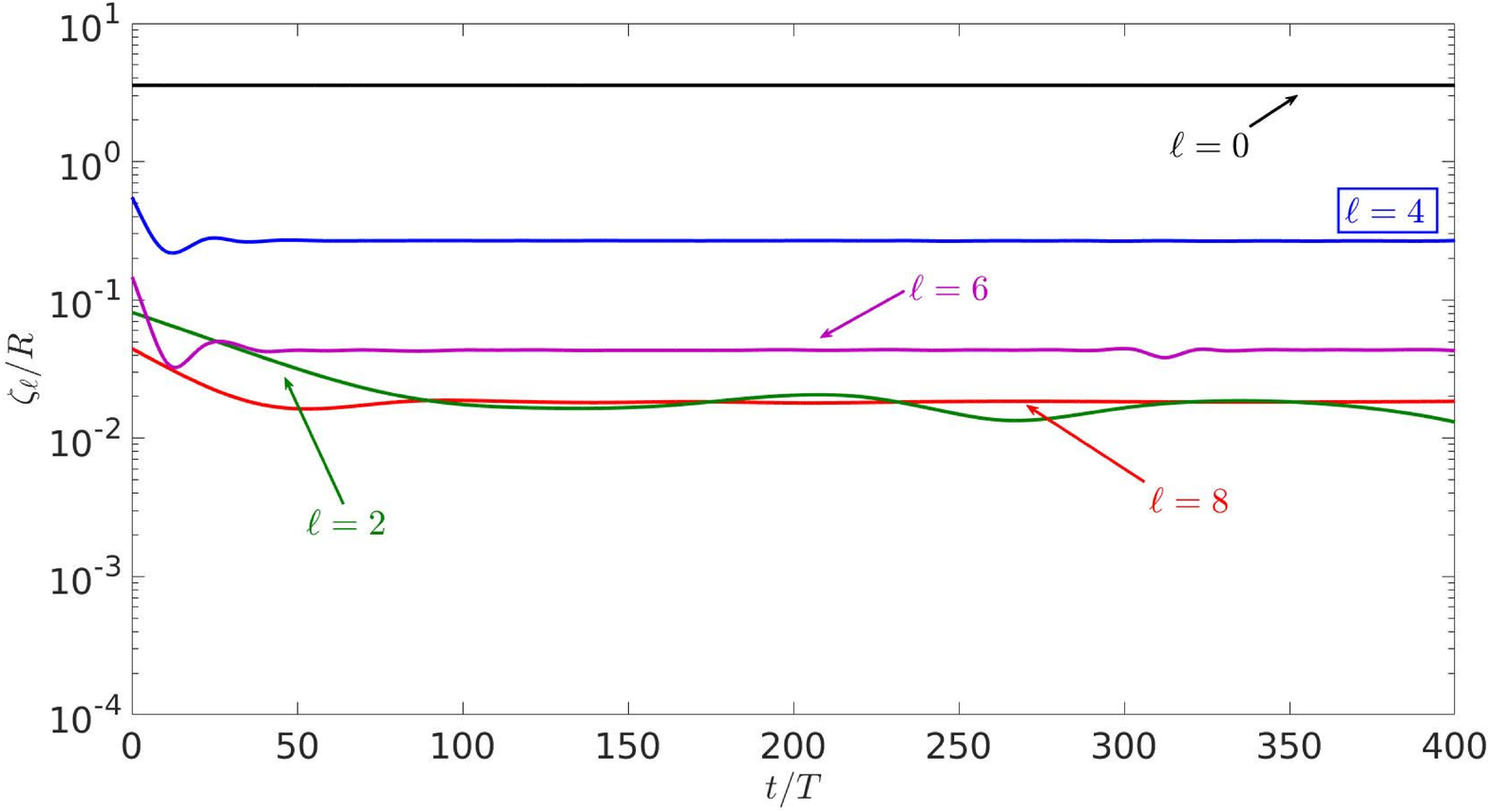}
\caption{Timeseries of $\ell$ components for a pattern whose
    dominant mode is $\lonset=4$ 
   when the initial  condition is cubic. The spectrum also
  contains important $\ell=2$ and $\ell=6=4+2$ components.}
\label{fig:S_ell4_L_cube}
\includegraphics[width=0.9\textwidth]{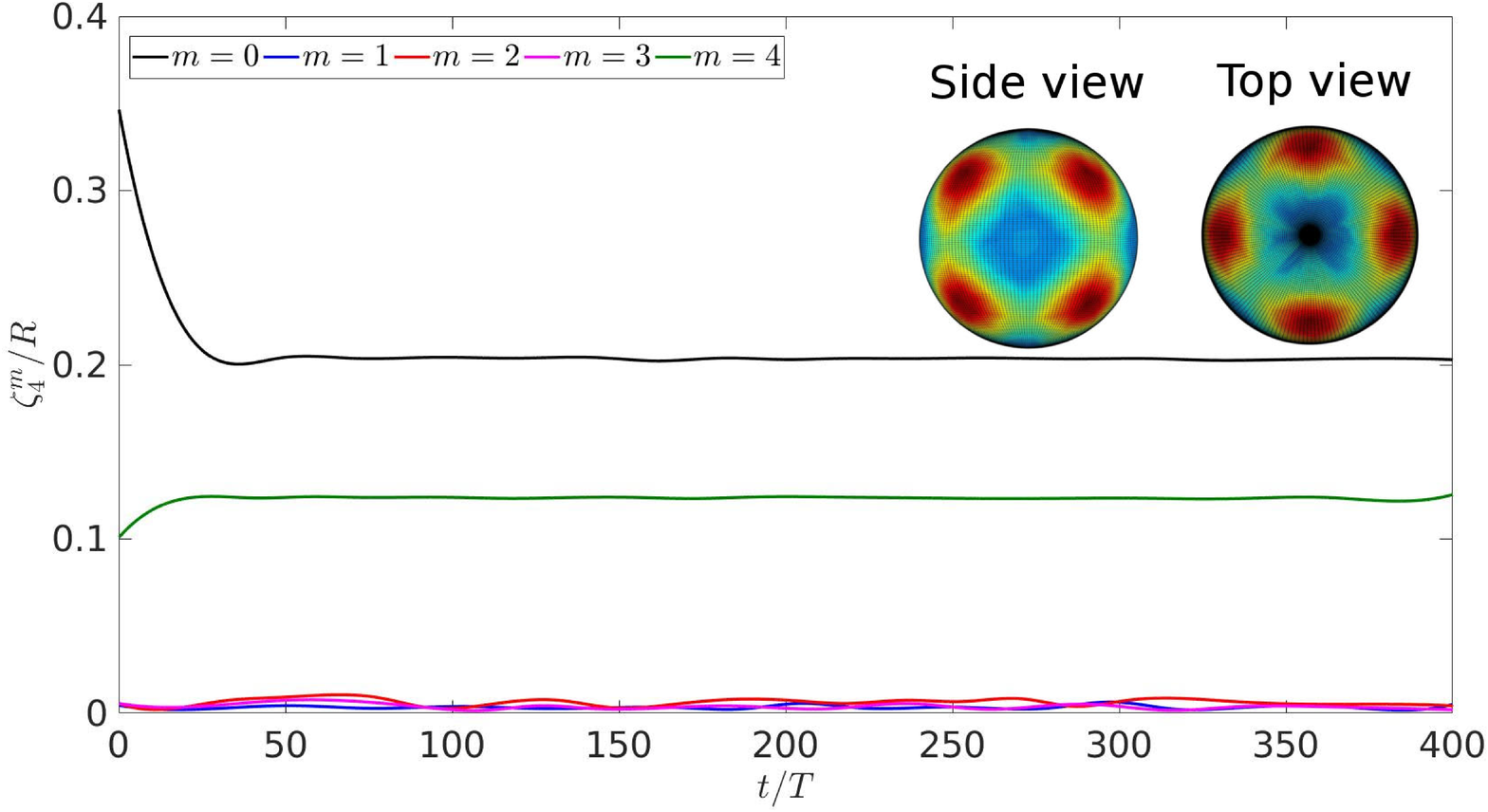}
\caption{Timeseries of $m$ components for an $\ell=4$ pattern
  when the initial condition is cubic. The pattern is stable 
 and consists of a superposition of modes $m=0$ and $m=4$, i.e. it is 
  cubic and aligned with the $z$ axis }
\label{fig:S_ell4_m_cube}
\end{figure}
\begin{figure}
\includegraphics[width=\textwidth]{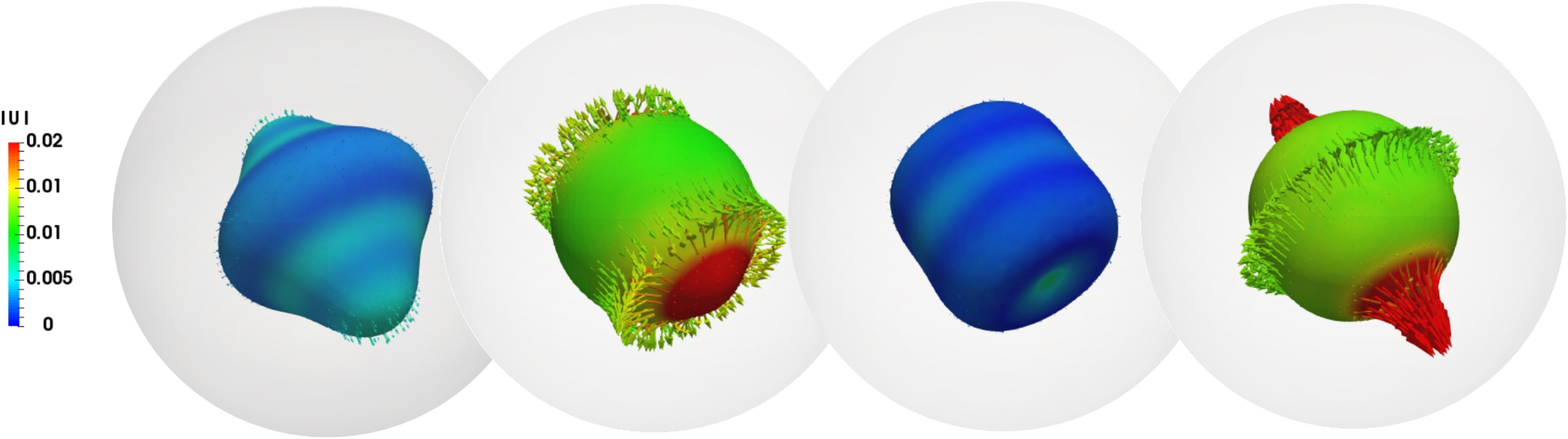}
\caption{Drop interface and velocity field for $\lonset=4$ axisymmetric pattern seen
  in subharmonic capillary waves at various phases during one reponse period $2T$.  The
  interface oscillates between a cylinder and a top-shaped object.
  Each drop is shown inside its spherical domain.
  Colors indicate the magnitude of the
  velocity (in m/s), which is maximal (minimal) when the surface is least
  (most) deformed.  Only outward-pointing velocity vectors are shown.
See also supplementary movie 4b.}
    \label{fig:l=4_axi}
\includegraphics[width=0.9\textwidth]{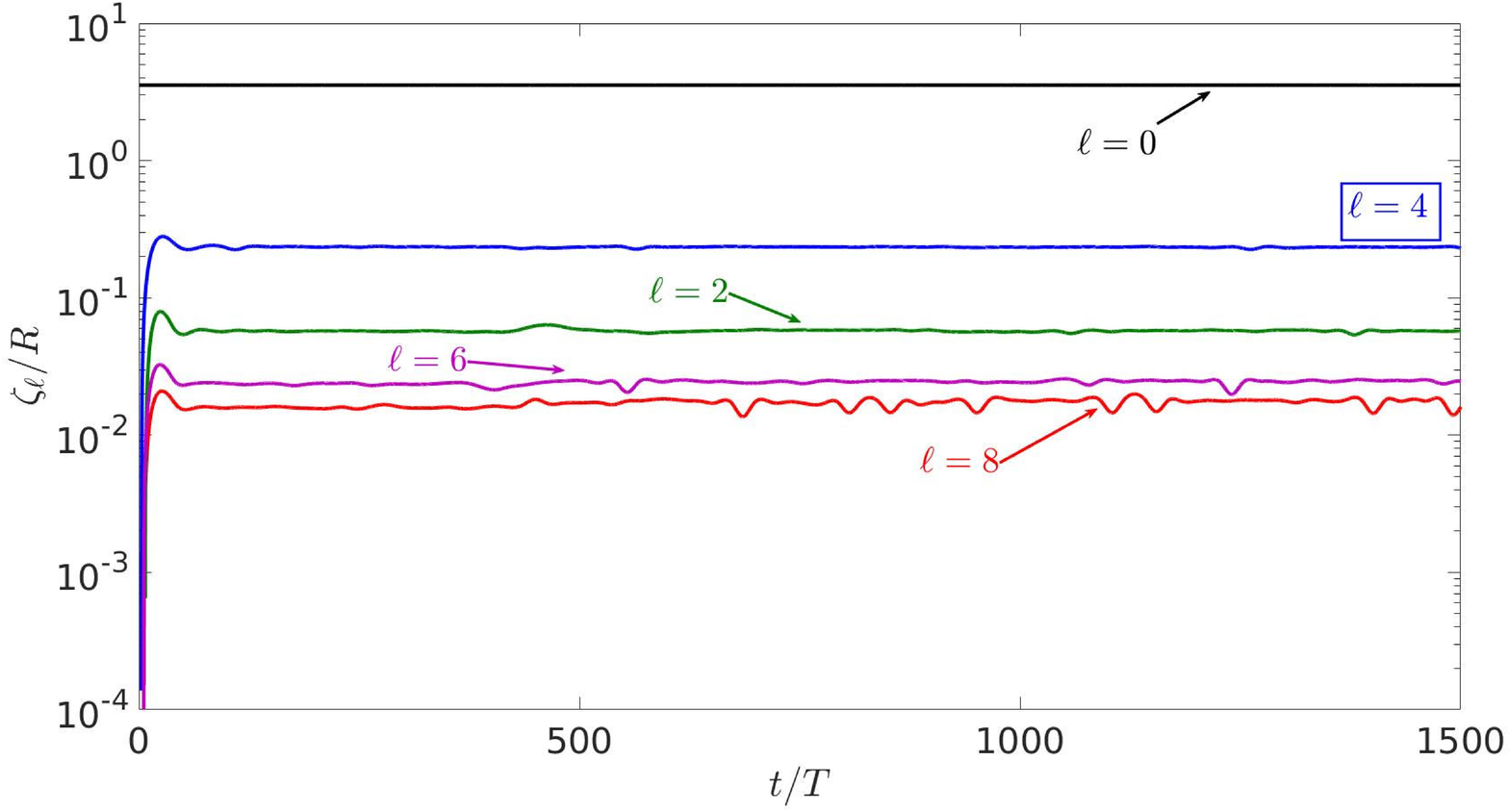}
\caption{Time evolution of amplitude of different $\ell$ components
  when the dominant wavenumber is $\lonset=4$ and the initial
  condition is axisymmetric. The spectrum also
  contains important $\ell=2$ and $\ell=6=4+2$ components.}
\label{fig:S_ell4_L_axi}
\includegraphics[width=0.9\textwidth]{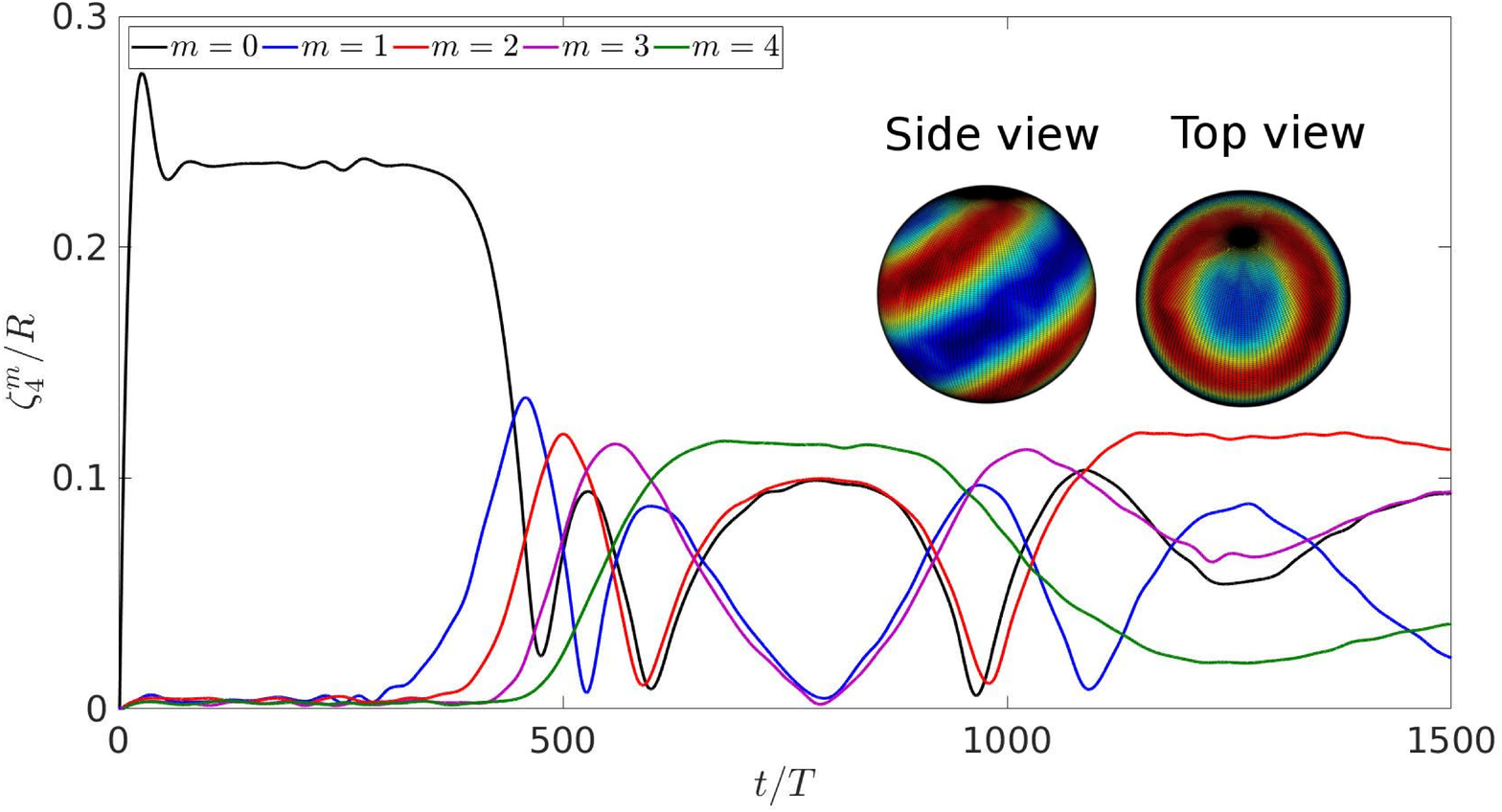}
\caption{Timeseries of $m$ components for an $\ell=4$ pattern
  when the initial condition is axisymmetric.
  The pattern remains axisymmetric, but its axis 
changes its orientation in time.}
\label{fig:S_ell4_m_axi}
\end{figure}

When $\lonset=4$, 
bifurcation theory predicts two possible solutions at onset
\citep{Busse1975,Chossat1991}, with axisymmetric or cubic symmetry,
produced by a transcritical bifurcation.
Neither branch is stable at onset but the cubic solution
is the preferred one \citep{Busse1975,Matth2003}.
We have been able to produce patterns of both kinds by starting
with initial conditions which are axisymmetric or cubic
\begin{equation}
\zeta - \Rd \propto \sqrt{7}\: Y_4^0 + \sqrt{5}\:Y_4^4 + {\rm c.c.}
\end{equation}

Visualisations of our numerical simulations of subharmonic capillary
oscillations with cubic symmetry are shown in figure \ref{fig:l=4}
and supplementary movie 4a.
The pattern oscillates between resembling 
a cube, with six square faces, of which three meet at each vertex, 
and its dual, the octahedron
with eight equilateral triangles, four of which meet at each corner.

Figure \ref{fig:S_ell4_L_cube} shows that, in addition to $\lonset=4$ and its
harmonics, the $\ell$ wavenumber spectrum contains a large $\ell=6$
component, as well as modes resulting from
interactions between the $\ell=4$ and $\ell=6$ families.  
The pattern is aligned with the numerical
domain, with two faces or minima (for the cube) or two vertices or
maxima (for the octahedron) located at the north and south poles.
With this alignment, the cubic solution
is expected to be a sum of $Y_4^0$ and $Y_4^{\pm 4}$.
Indeed figure \ref{fig:S_ell4_m_cube} shows that $\zeta_4^0$ and $\zeta_4^4$
are dominant and constant.
We have also obtained a cubic pattern in simulations (not presented here)
of harmonic gravitational waves with a perturbed spherical initial
condition and a resolution of $256^3$. 

From an axisymmetric initial condition, the solution remains
axisymmetric over the time of our simulation, but develops 
a tilting and rotating axis of symmetry, as was the case for $\ell=2$.
Visualisations of this state are shown in figure \ref{fig:l=4_axi}
and supplementary movie 4b.
Figure \ref{fig:S_ell4_L_axi} shows the evolution of the amplitudes
of the various $\ell$ modes. We observe that 
$\zeta_2$ is higher (a feature we have observed whenever
a pattern is axisymmetric) and $\zeta_6$ is lower
than was the case for the cubic solution.
The oscillatory behavior of
the various $m$ modes in figure \ref{fig:S_ell4_m_axi}
are manifestations of the fact that the axis of symmetry
rotates, as in figure \ref{fig:S_emm_L2}.




\subsection{Case $\lonset = 5$}
\begin{figure}
\includegraphics[width=\textwidth]{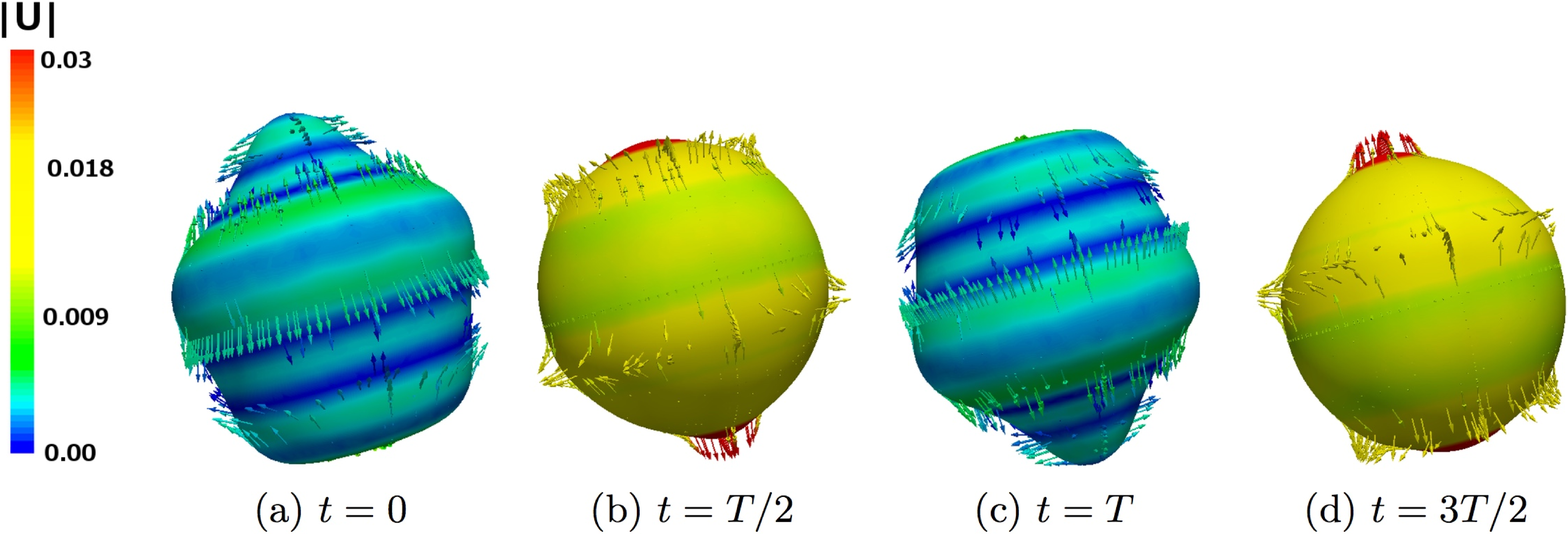}
\caption{Drop interface and corresponding velocity field for transient
  axisymmetric $\lonset=5$ pattern seen in subharmonic capillary waves
  over one reponse period $2T$.  Colors indicate the magnitude of the
  velocity (in m/s), which is maximal (minimal) when the surface is least
  (most) deformed.  Only outward-pointing velocity vectors are shown.
See also supplementary movie 5.}
    \label{fig:l=5}
\includegraphics[width=\textwidth]{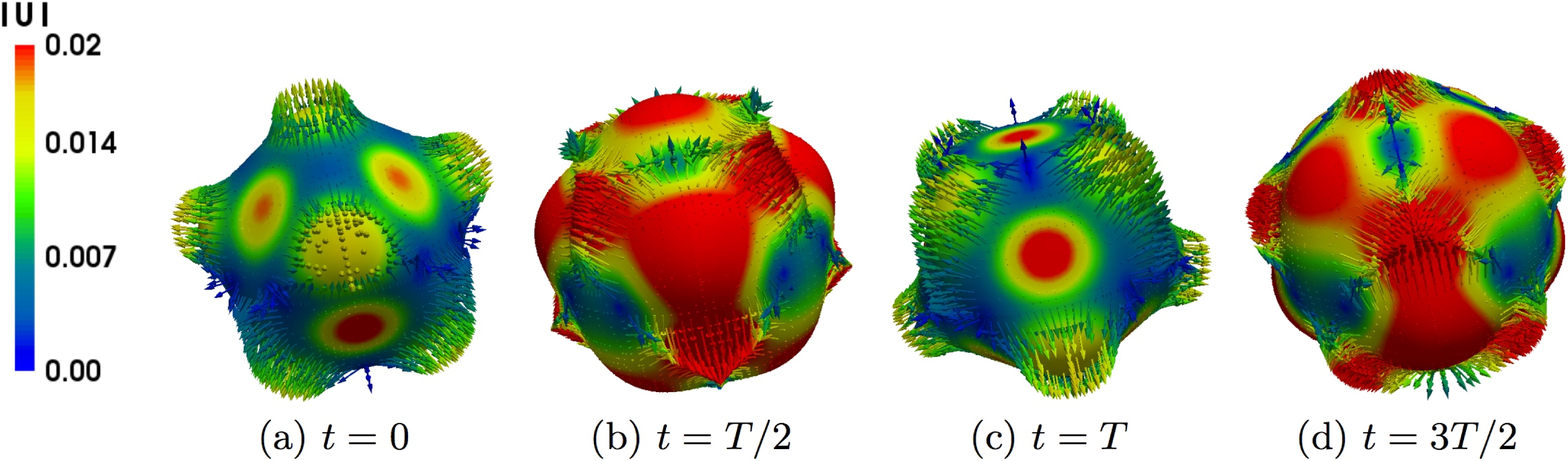}
\caption{Drop interface and corresponding velocity field for final $\lonset=5$
  pattern with $D_4$ symmetry.
See also supplementary movie 5.}
    \label{fig:l=5_D4}
\end{figure}

\begin{figure}
\centering
\includegraphics[width=0.9\textwidth]{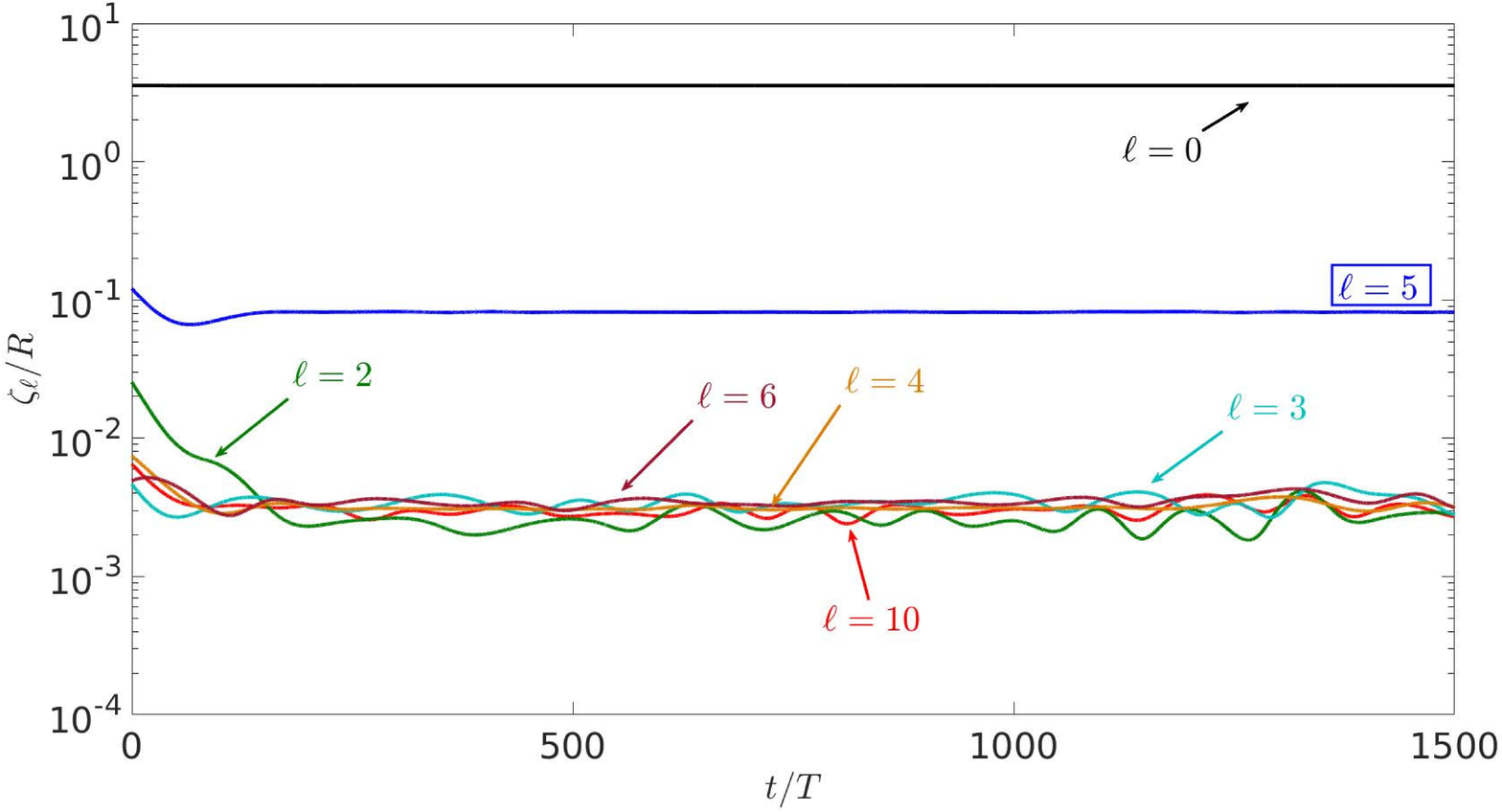}
\caption{Timeseries of $\ell$ components for a pattern whose
    dominant mode is $\lonset=5$ 
  when the initial condition is axisymmetric.
  Many components are present, in addition to the expected multiples of 5.}
    \label{fig:S_L5}
\includegraphics[width=0.9\textwidth]{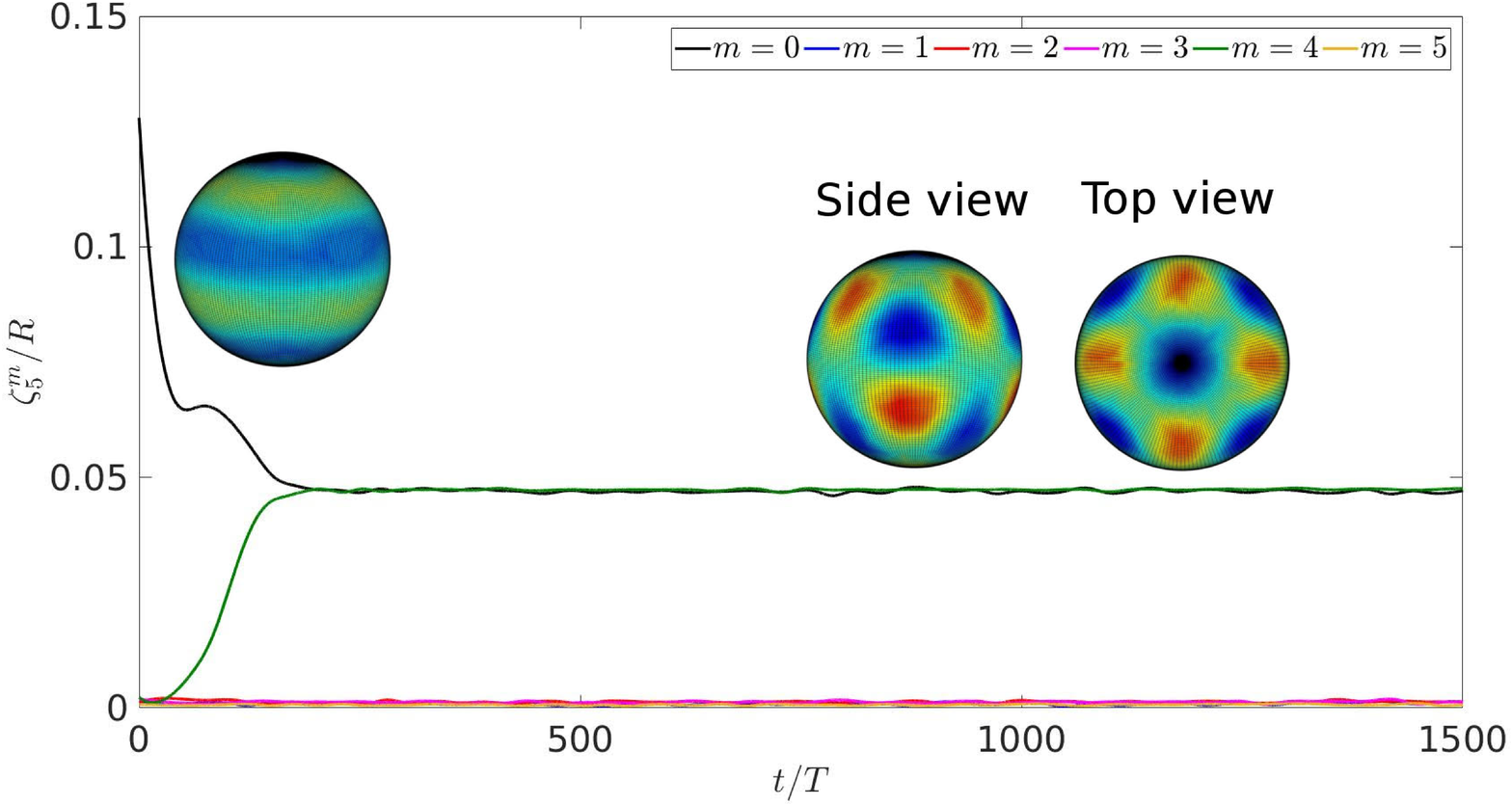}
\caption{Timeseries of $m$ components for an $\ell=5$ pattern.
  An initially axisymmetric pattern with $m=0$
  briefly equilibrates and is then replaced by a $D_4$ pattern, which is a
  superposition of $m=0$ and $m=4$.}
    \label{fig:S_emm_L5}
\end{figure}

The theoretical analysis of the $\lonset=5$ case is the most complicated
of those surveyed here \citep{BR1982,Riahi1984},
since it leads to eight allowed solutions with different
symmetries \citep{Chossat1991}. We find two of these solutions
in our simulations, one as a short-lived unstable equilibrium
and the other as a long-term asymptotic state.

Starting from an initial condition which
is an axisymmetric perturbation of the sphere, 
the solution quickly equilibrates to axisymmetric oscillations
which are shown in figure \ref{fig:l=5}.
The solution is then  replaced by another solution with $D_4$ symmetry,
shown in figure \ref{fig:l=5_D4}.
See supplementary movie 5.
In the $D_4$ solution, the upper hemisphere contains four patches
of outward (inward) flow, and the lower hemispere contains four
similar patches, located at longitudes which are halfway between
those of the patches of the upper hemisphere.
This pattern corresponds to the preferred solution called $\mathbf{F}$ by \cite{Riahi1984}:
\begin{equation}
\zeta - \Rd \propto \sqrt{3}\: Y_5^0 + \sqrt{5} \: Y_5^4 + {\rm c.c.}
\label{eq:Riahi}  \end{equation}

The spectra in $\ell$ and $m$ are shown in figures \ref{fig:S_L5}
and \ref{fig:S_emm_L5}.
Because the pattern remains aligned with the $z$ axis,
the transition from the axisymmetric pattern to \eqref{eq:Riahi}
can be tracked by following the different $m$ modes, 
as is done in figure \ref{fig:S_emm_L5}.
The initial plateau in $\zeta_5^0$ 
indicates that the axisymmetric phase of the oscillations shown in 
figure \ref{fig:l=5} comprises a solution, albeit unstable, of our
system. (This is accompanied in figure \ref{fig:S_L5} by a brief slowing of
the decay of the $\ell=2$ mode which is present in all of our
nonlinear axisymmetric solutions.)
The subsequent decrease in $\zeta_5^0$ is accompanied by an
increase in $\zeta_5^4$, which together comprise the stable solution
with $D_4$ symmetry.



\subsection{Case $\lonset = 6$}
\label{sec:ell=6}

\begin{figure}
\includegraphics[width=\textwidth]{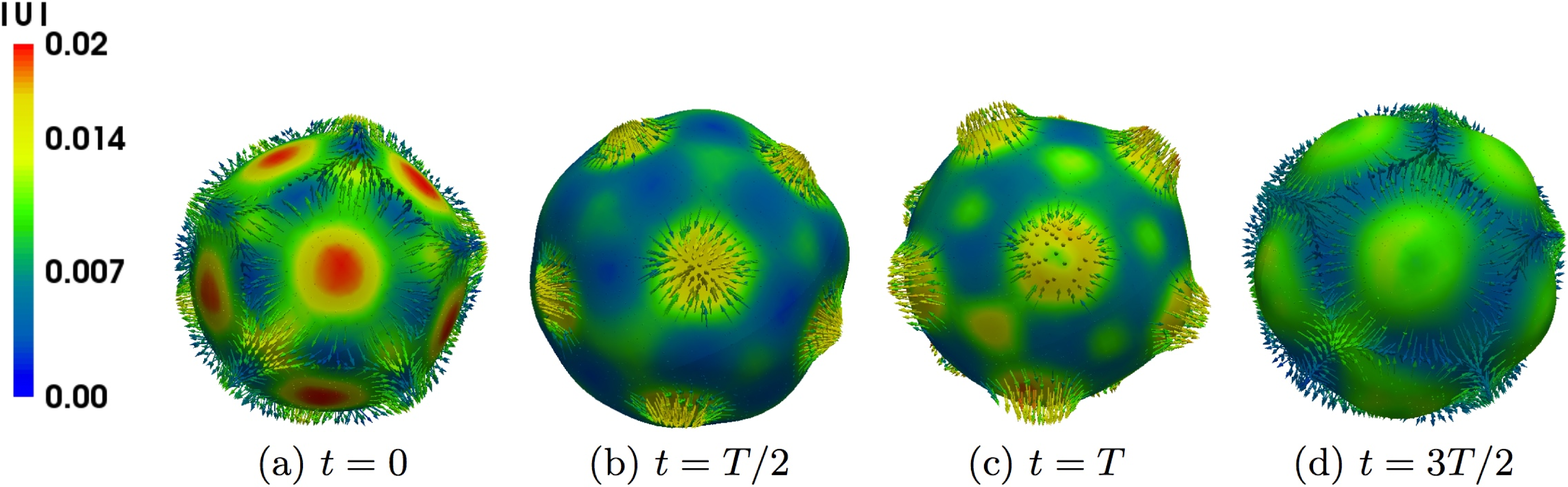}
 \caption{Drop interface and corresponding velocity field for $\lonset=6$
   transient pattern with icosahedral symmetry of subharmonic capillary
   waves over one reponse period $2T$.  Colors indicate the magnitude
   of the velocity (in m/s), which is maximal (minimal) when the surface is
   least (most) deformed.  Only outward-pointing velocity vectors are
   shown. See also supplementary movie 6.}
    \label{fig:l=6_icos}
\includegraphics[width=\textwidth]{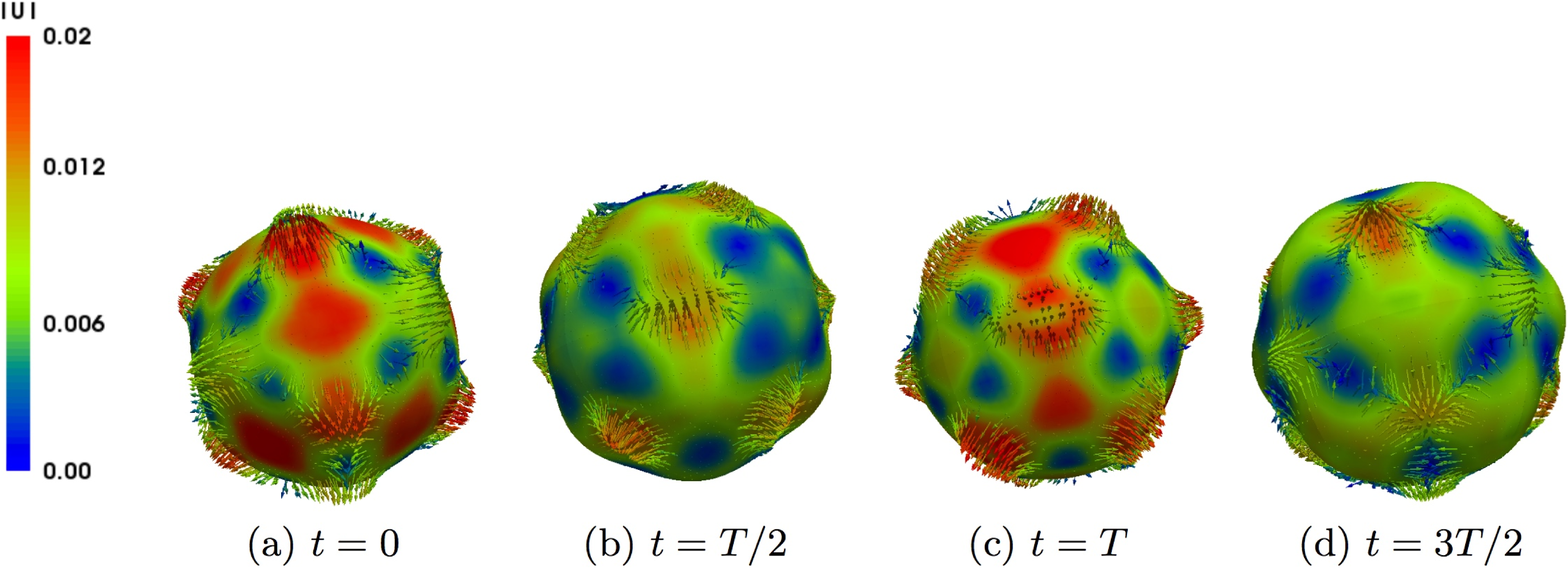}
\caption{Drop interface and corresponding velocity field for
  $\lonset=6$ transient pattern with cubic symmetry.}
    \label{fig:l=6_cubic}
\end{figure}

\begin{figure}
\centering
\includegraphics[width=0.9\textwidth]{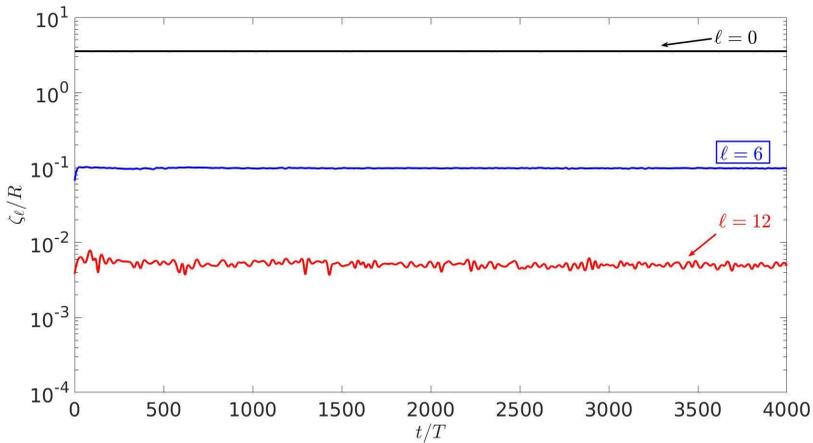}
\caption{Time evolution of the amplitudes of modes with different
  $\ell$ values when the dominant mode is $\lonset=6$.}
    \label{fig:S_L6}
\end{figure}
\begin{figure}
\centering
\includegraphics[width=\textwidth]{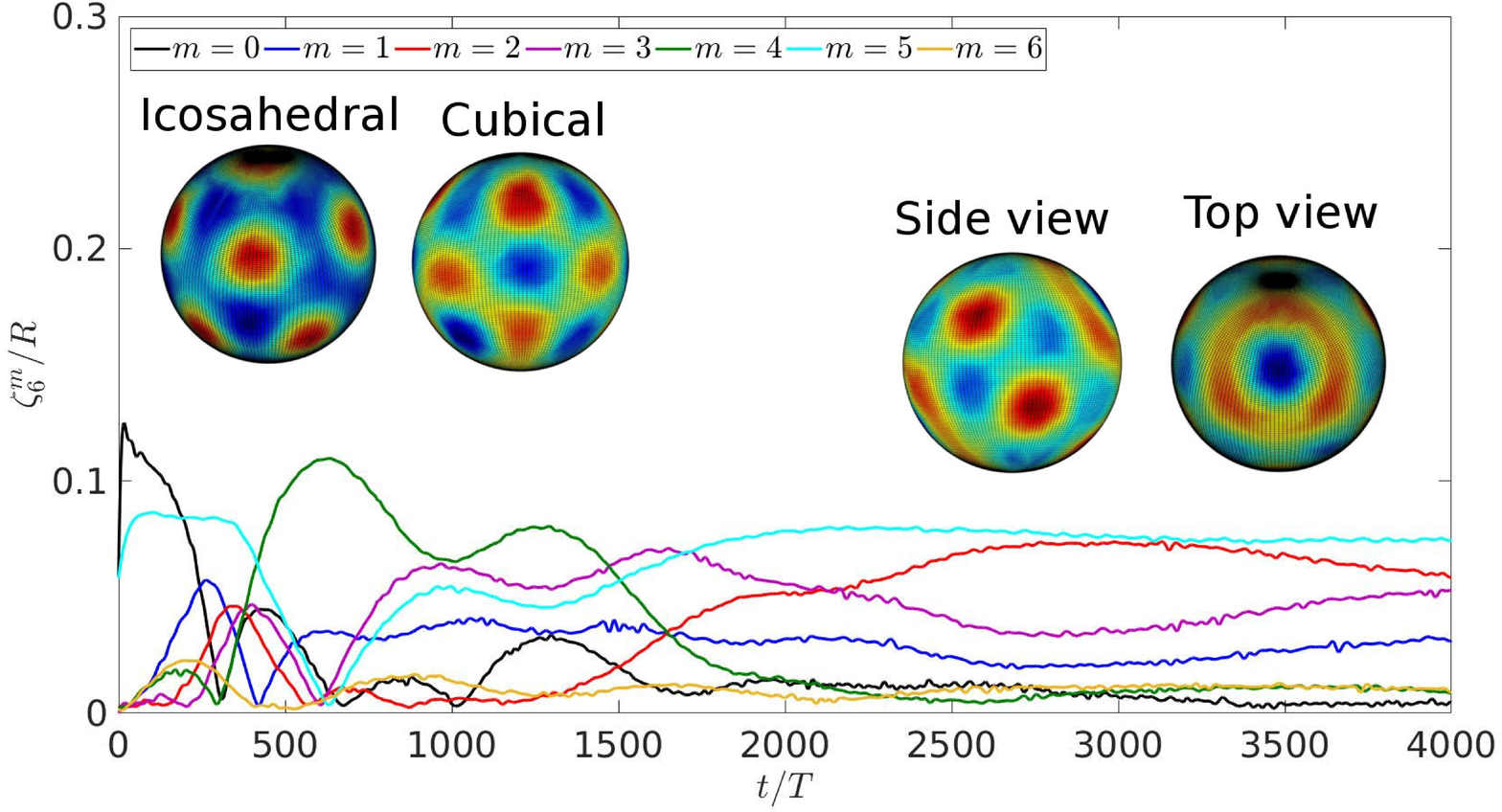}
\caption{Timeseries of $m$ components for an $\ell=6$ pattern.
  From an icosahedral initial condition composed of modes
  $m=0$ and $m=5$, the pattern tilts away from the $z$ axis and by
  $t/T\approx 300$ acquires cubic symmetry with dominant
  mode $m=4$. The subsequent increase in mode $m=3$ at $t/T\approx 800$
  is associated with tilting of orientation of the cubical pattern
  away from the $z$ axis. Insets show projections of
  the initial icosahedral pattern, the subsequent cubic pattern,
  and the solution at later times, which is neither of these.}
    \label{fig:S_emm_L6}
\end{figure}

We compute subharmonic capillary oscillations for $\ell=6$.
For $\lonset=6$, there exist four possible solutions:
axisymmetric, six-fold symmetric ($D_6$) octahedral, 
and icosahedral. Of these, the icosahedral solution is 
preferred near threshold, via a maximization argument \citep{Busse1975}
or a stability argument \citep{Matth2003}.
With this in mind, we use as an initial condition an
icosahedral perturbation of the sphere \citep{Busse1975}:
\begin{equation}
\zeta - \Rd \propto \sqrt{11} \: Y_6^0  + \sqrt{14} \: Y_6^5 + {\rm c.c.}
\end{equation}
Figure \ref{fig:l=6_icos} shows the icosahedral/dodecahedral oscillations
during the first phase of the simulation, with a clear
five-fold symmetry; see also supplementary movie 6.
Figure \ref{fig:l=6_icos}(a) resembles a dodecahedron, i.e. 
pentagons which meet in sets of three at the vertices,
while \ref{fig:l=6_icos}(c) resembles an icosahedron, i.e. 
triangles which meet in sets of five at the vertices. 
Further evolution leads to a second phase of the solution,
shown in figure \ref{fig:l=6_cubic}, 
which oscillates between a cube and its dual, an octahedron.
Indeed, simulations from a spherical initial condition (perturbed by its discrete
representation on a triangular mesh) also led directly to the cubic/octahedral solution.

Figure \ref{fig:S_L6} shows the time evolution of the spectrum in $\ell$,
while figure \ref{fig:S_emm_L6} shows the time evolution of
the most important $m$ modes associated with $\ell=6$.  The icosahedral pattern,
consisting of modes $m=0$ and $m=5$, quickly tilts away from the $z$
axis and by $t/T\approx 300$ is replaced by a cubic pattern, whose dominant mode is
$m=4$. The subsequent increase in $m=3$ is associated with the
tilting of the cubic pattern away from the $z$ axis.
Eventually, however, the solution shifts to the pattern shown in the
rightmost insets of figure \ref{fig:S_emm_L6}, 
which is neither icosahedral nor cubic.



\section{Discussion}

Using the parallel front-tracking code {\tt BLUE}, 
we have been able to simulate the spherical Faraday problem 
and thereby to produce patterns with all spherical harmonic
wavenumbers between $\lonset=1$ and $\lonset=6$.
We have simulated both gravitational and capillary waves,
in both the harmonic regime and the more usual subharmonic regime.


Our simulations agree in most cases with two types of theory.
First, the spherical harmonic wavenumber $\lonset$ obtained
in each case agrees with the results of Floquet analysis, presented
in \cite{Ali1} and in figure \ref{fig:BLUE_tongues}.
Second, the interface shapes we observe are readily interpreted using
the theory of pattern formation on the sphere
\citep{Busse1975,BR1982,Riahi1984,IG1984,Golubitsky,Chossat1991,Matth2003}.
For $\lonset=1$ and
$\lonset=2$, only one type of pattern is possible, and that is the one we
observe. For our vibrating drop, the $\lonset=1$ ``pattern'' is
manifested as a back-and-forth motion of the spherical drop, while the
$\lonset=2$ pattern alternates between an oblate and a prolate spheroid.
For $\lonset=3$, we observe a tetrahedral pattern, one of the two
solutions predicted to be stable (out of the three which can exist),
while for $\lonset=4$, depending on the initial conditions, 
we observe both of the possible solutions, 
a cubic/octahedral or an axisymmetric pattern. For $\lonset=5$, we observe an
unstable axisymmetric and a stable $D_4$ pattern.  These are
two of the eight possible solutions; $D_4$ is one of those which can
be stable at onset.  For $\lonset=6$, an initially icosahedral
solution makes a transition to a cubic/octahedral pattern,
which is succeeded by a solution with neither symmetry.
It is surprising to find such good agreement with theory
given the differences with our configuration mentioned in the introduction,
i.e. the fact that our patterns are oscillatory rather than steady and
that our parameters are far above threshold.  
A complete study of each of these cases, varying the forcing amplitude
from threshold to higher values, would be desirable.
Another crucial issue, both empirical and mathematical, is the possible difference between
the harmonic and subharmonic regimes in each of the cases.

An important avenue of exploration that has arisen in our study
is that of the long-term dynamics.
The $\ell=2$ (prolate-oblate) and $\ell=4$ (axisymmetric)
show the drop tumbling into and out of alignment with the coordinate system, 
as illustrated in figures \ref{fig:S_emm_L2} and \ref{fig:S_ell4_m_axi}.
The $\ell=3$ (tetrahedral) and $\ell=6$ (cubic/octahedral) oscillations
in figures \ref{fig:S_emm_L3} and \ref{fig:S_emm_L6}
also show very long phases (2000 or more forcing periods $T$) during
which the orientation continues to change.
Another case in which very long-term Faraday-wave dynamics
has been found is that of hexagonal waves in a minimal
domain \citep{Perinet2012}.
To the best of our knowledge, there exists no 
explanation of such long-term dynamics 
in terms of pattern formation, fluid dynamics, or any other kind of theory.



We have demonstrated the feasibility of well-resolved numerical
simulation of the spherical version of the Faraday instability over
extremely long times.  We believe that the Faraday problem serves as a
rigorous proving ground for numerical interface techniques. The wide
variety of drop shapes and their detailed patterns simulated here
demonstrate the necessity and advantages of using high fidelity
numerical techniques developed for accurately computing two-phase
flows and free-surfaces particularly where precise volume
conservation, interface advection and calculation of surface tension
forces are fundamental. We hope that this work on the spherical
Faraday instability can open up new possibilities in the study of
pattern formation and that the numerical code can serve as a useful
tool in the exploration of this rich dynamical system.  In addition,
we believe that the results obtained here serve both to validate the
techniques implemented in BLUE and also to provide encouragement in
applying the code to two-phase flow scenarios in highly non-linear regimes.

\subsubsection*{Acknowledgements}
This work was performed using high performance computing resources provided by 
the Institut du Developpement
et des Ressources en Informatique Scientifique (IDRIS) of the Centre
National de la Recherche Scientifique (CNRS), coordinated by
GENCI (Grand Equipement National de Calcul Intensif) through
grants A0042B06721 and A0042A01119.
L.S.T. acknowledges support from the Agence Nationale de la Recherche (ANR)
through the TRANSFLOW project.
This research was supported by Basic Science Research Program through the
National Research Foundation of Korea (NRF) funded by the Ministry of Education
(No.~2017R1D1A1B03028518).

\newpage
\bibliographystyle{jfm}
\bibliography{sphere}


\end{document}